\documentclass[a4paper,11pt]{article}
\usepackage{jinstpub} % for details on the use of the package, please see the JINST-author-manual
\usepackage{lineno}
\usepackage{subfigure,dcolumn}
\usepackage{epstopdf}
\usepackage{mhchem}
\usepackage{upgreek}
\usepackage{graphicx}
\usepackage{xcolor}
\usepackage{amsmath}
\usepackage{ulem}
\usepackage{multirow}
\title{\boldmath Dark Count of 20-inch PMTs Generated by Natural Radioactivity}

\emailAdd{wangzhm@ihep.ac.cn}
\author[a,b]{Yu Zhang}
\author[a,b,c]{Zhimin Wang}
\author[a,b]{Min Li}
\author[a,b]{Caimei Liu}
\author[a,b]{Narongkiat Rodphai}
\author[a,b,c]{Yongpeng Zhang}
\author[a,b,c]{Jilei Xu}
\author[a,b]{Changgen Yang}
\author[a,b,c]{Yuekun Heng}
\affiliation[a]{Institute of High Energy Physics, Beijing 100049, China}
\affiliation[b]{University of Chinese Academy of Sciences, Beijing 100049, China}
\affiliation[c]{State Key Laboratory of Particle Detection and Electronics, Beijing 100049, China}

\abstract{The primary objective of the JUNO experiment is to determine the ordering of neutrino masses using a 20-kton liquid-scintillator detector. The 20-inch photomultiplier tube (PMT) plays a crucial role in achieving excellent energy resolution of at least 3\,\% at 1\,MeV. Understanding the characteristics and features of the PMT is vital for comprehending the detector's performance, particularly regarding the occurrence of large pulses in PMT dark counts. This research paper aims to further investigate the origin of these large pulses in the 20-inch PMT dark count through measurements and simulations. Our results confirm that natural radioactivity and muons striking the PMT glass are the main sources of the large pulses. We evaluate their contribution quantitatively by performing spectrum fitting. By analyzing the PMT dark count spectrum, it becomes possible to roughly estimate the radioactivity levels in the surrounding environment.}


\arxivnumber{2307.15104} % Only if you have one

\begin{document}
\maketitle
\flushbottom

\section{Introduction}
\label{1:intro}
Photomultiplier tubes (PMTs) are extensively utilized in particle physics experiments that require single-photon-sensitive light detection, such as Super-K\,\cite{Super-Kamiokande:1998uiq,PhysRevD.83.052010}, KamLAND\,\cite{PhysRevLett.90.021802}, SNO\,\cite{SNO}, IceCube\,\cite{PhysRevLett.110.131302}, Double Chooz\,\cite{chooz}, Daya Bay\,\cite{dayabay}, and RENO\,\cite{KIM201324}. The performance and characterization of PMTs have been extensively studied and well-understood\,\cite{AugerPMT,GE2016175,BorexinoPMT,DayabayPMT,ChoozPMT,HKPMT,JUNO3inchPMT,JUNOPMTinstr,KM3NeTPMT,JUNOPMTflasher,MCPPMT2018,YWang_newMCP,wavesamplingPMT,waveAnalysisHaiqiong}.

\par However, recent rare-event neutrino experiments worldwide, such as Double Chooz\,\cite{Abe_2016}, Daya Bay\,\cite{DWYER201330}, IceCube\,\cite{IceCube-inproceedings}, and RENO\,\cite{JANG2014145}, have encountered issues caused by large pulses in PMTs. It is known that the passage of a muon through the PMT glass (e.g., Hamamatsu PMT R5912) generates a large pulse due to Cherenkov radiation, as studied in\,\cite{PMTmuon2007,BAYAT20141,Zhang_2022}. Additionally, ongoing research is being conducted on scintillation glass\,\cite{glass-2015,TANG2022112585-2022,AMELINA2022121393-2022}. %Studies have also focused on flasher (the PMT internal self-generated light) \cite{CAO201662-dayabay-flasher,Qian_2020,JUNOPMTflasher}, after-pulse related big pulses \cite{MIRZOYAN199774-AP,JUNO-PMT-AP} and large pulses, as seen in\,\cite{Yang_2020}. 
Studies have also focused on flasher (unlike genuine light emission induced by dynodes, see in \cite{7104168}, also there is light emission caused by a discharge on the sharp edge inside the PMT or the voltage divider) \cite{CAO201662-dayabay-flasher,Qian_2020,JUNOPMTflasher}, after-pulse related big pulses \cite{MIRZOYAN199774-AP,JUNO-PMT-AP} and large pulses, as seen in \,\cite{Yang_2020}.
To expand on previous investigations regarding muon-generated large pulses, we further explore the influence of natural radioactivity for a comprehensive understanding of related signals from PMTs.

\par The Jiangmen Underground Neutrino Observatory (JUNO)\,\cite{JUNOCDR,JUNOphysics} is currently under construction in Jiangmen, Guangdong, China. JUNO aims to study neutrino mass ordering with 3\% energy resolution at 1\,MeV, accurately determine neutrino oscillation parameters, and explore other aspects of neutrino physics using a 20-kton liquid scintillator monitored by up to 20,000 high quantum efficiency (QE) 20-inch PMTs. JUNO has selected two types of 20-inch PMTs\,\cite{JUNOdetector,JUNOPMTinstr}: 5,000 Hamamatsu Photonics K.K. (HPK, R12860) dynode PMTs\,\cite{HPK-R12860} and 15,000 newly developed microchannel plate (MCP) PMTs from North Night Vision Technology (NNVT) Co., LTD (MCP, GDB6201)\,\cite{NNVT-GDB6201-note}.

\par In this article, we present a detailed investigation of the dark count rate (DCR) of 20-inch PMTs associated with natural radioactivity. Section\,\ref{1:largepulse} provides a brief description of the testing system and configurations. A dedicated simulation is performed for better understanding and is compared with the experimental measurements, as presented in section\,\ref{1:sim}. The measurement results are shown in section\,\ref{1:comp}.

\section{Large Pulses from PMT Dark Count}
\label{1:largepulse}

The dark count in a photomultiplier tube (PMT) primarily arises from the thermal electron emission of the PMT photocathode when the PMT is in a dark environment. In general, the amplitude of dark counts is expected to be at the level of single photoelectron (SPE), mostly below 3\,p.e.~(photoelectrons), and the rate is much lower than 1\,Hz for signals larger than 3\,p.e.~(assuming DCR in SPE from photocathode approximately 10\,kHz and a 10\,ns coincidence window) \cite{HamManual,POLYAKOV201369} and the rate of big pulses related to after pulse in dark is also much lower than 1\,Hz \cite{MIRZOYAN199774-AP,JUNO-PMT-AP}. For a better understanding of the sources contributing to the dark count, including thermal emission, flashers of the 20-inch PMT, after pulse, muons, and natural radioactivity, a 20-inch PMT test system is employed to measure the dark count rate (DCR) for large pulses \cite{zhang2022study}. This measurement aims to investigate the DCR versus threshold, as well as the amplitude and charge spectra of the dark count.

\par
\begin{figure*}[!hbtp]
	\centering
	\subfigure[HPK PMT]{
        \includegraphics[width=0.45\hsize]{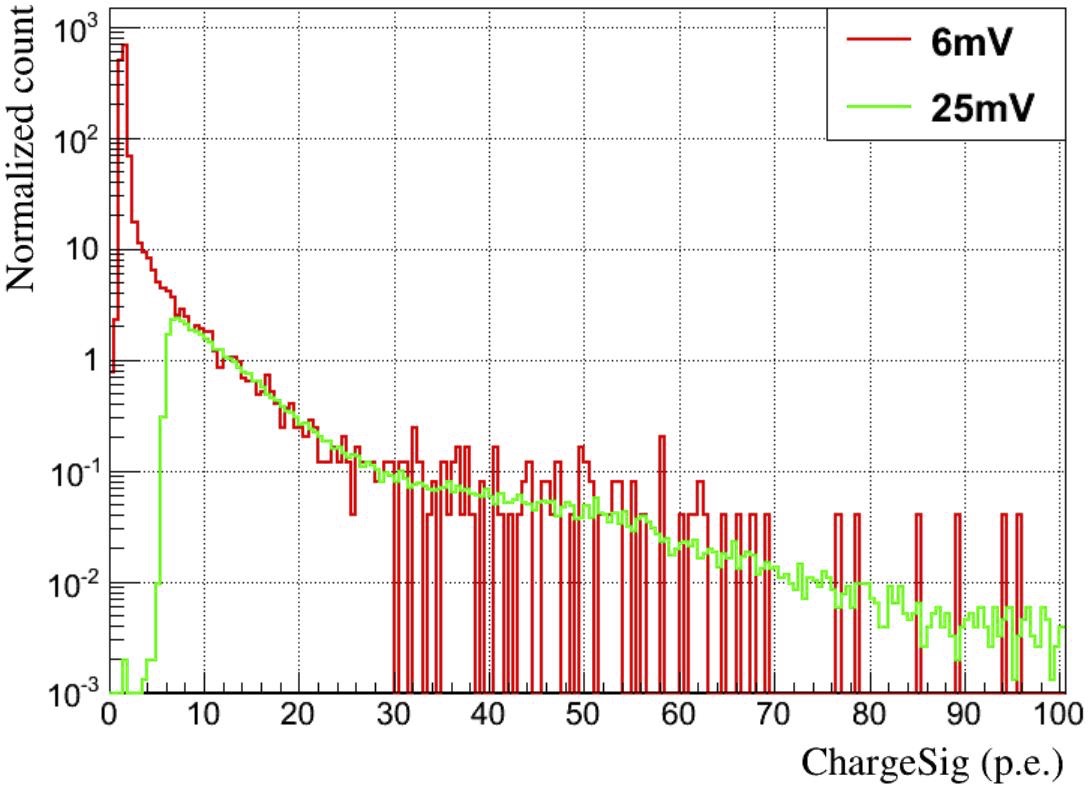}
        \label{fig: dn-amp-hpk}
    }
    \quad
	\subfigure[NNVT PMT]{
        \includegraphics[width=0.45\hsize]{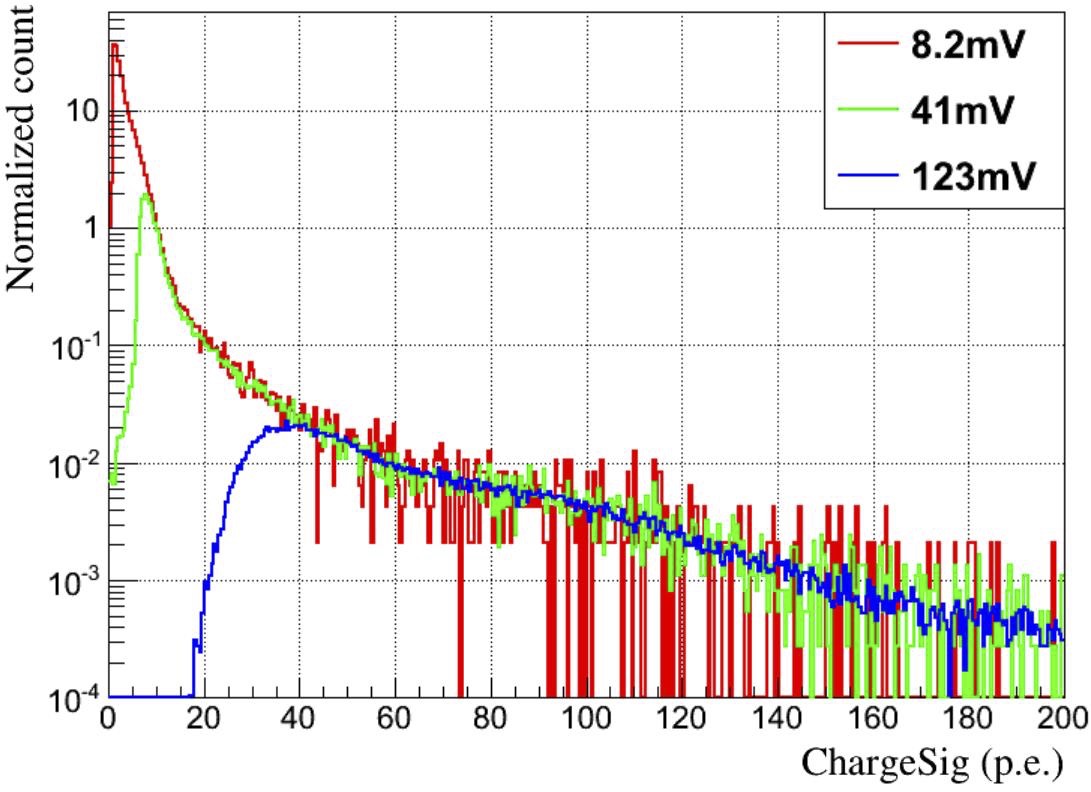}
        \label{fig: dn-amp}
    }   
    \caption{The normalized charge spectra of 20-inch PMTs' dark count measured with different thresholds on the signal amplitude by a discriminator are presented in this study. (a) shows the charge spectrum of the 20-inch HPK PMT, while (b) displays the charge spectrum of the 20-inch NNVT PMT. Note: (1) The threshold set to a discriminator used in the measured is applied to the signal amplitude in mV (it is around 8\,mV per p.e.\,with a gain of $1 \times 10^{7}$), which will introduce a range rather than a sharp edge on the converted charge spectrum. (2) The NNVT PMT shows a longer tail in higher charge range than the HPK PMT, which will be further discussed later.}
    \label{fig:dn:wave}
\end{figure*}

 The dark count of the 20-inch PMT is extensively examined through detailed measurements. Waveforms of PMT pulses are captured during the threshold survey for both HPK and NNVT PMTs using a test system (the gain of both kind of PMTs is set to $1 \times 10^{7}$, where the typical amplitude of a single photoelectron (p.e.) signal is around 8\,mV). The charge distributions of the DCR are depicted in Figure \ref{fig:dn:wave} (calculated by a peak gain model as discussed in ref.\,\cite{JUNOPMTgain}), where all plots are normalized relative to the result obtained at the lowest threshold, considering only events with amplitudes higher than 25\,p.e. Both types of PMTs behave similar overall trends in their charge spectra, showing a distinctive structure. Charge levels from zero to around 3\,p.e.~are primarily contributed by thermal electron emission. The large signals ranging from approximately 20 p.e.~to at least 150\,p.e.~in charge are primarily attributed to cosmic muons \cite{zhang2022study}. On the other hand, signals ranging from approximately 3\,p.e.~to 20\,p.e.~in charge require further investigation and are in the focus of this study. 

\par
\begin{figure}[!hbt]
    \centering
    \includegraphics[width=0.45\hsize]{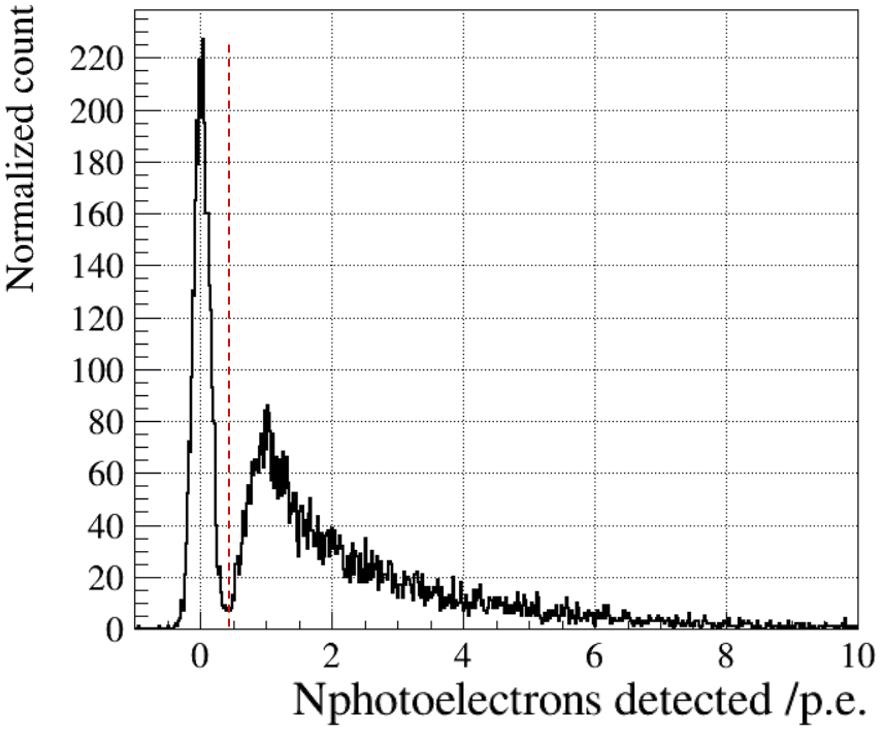}
    \caption{The single photoelectron (p.e.) spectrum of the NNVT PMT was measured using a pulse light source and external trigger (the mean value of the number of impinging photons is around 0.1)\,\cite{PMTgainmodel1994,Luo_2019}. Only events with amplitudes higher than 0.4 p.e.~in the spectrum (around the valley of the distribution) were considered for the smearing analysis to remove the non-signal entries. The long tail of the single p.e.~spectrum of the NNVT PMT could be related to the back-scattering of the electrons on the MCP\,\cite{MCPPMTTTSsen-reflection}.}
    \label{fig:spe}
\end{figure}

 In a simulation, the charge from the PMT is represented in individual photoelectrons (p.e.) before undergoing further processing, making it difficult to directly compare with real data. As illustrated in Figure\,\ref{fig:rad-hpk}, it is necessary to apply charge smearing (convert the ideal and individual photoelectron to measured value considering the real charge response of the PMTs with resolution) to the simulated charge before comparing it with the experimental measurements, particularly for the NNVT PMT, which exhibits a long tail in its charge spectrum\,\cite{JUNOPMTgain}. The long tail of the single p.e.~spectrum of the NNVT PMT could be related to the back-scattering of the electrons on the MCP\,\cite{MCPPMTTTSsen-reflection}, which should be a similar effect as the dynode PMTs\,\cite{PMTgainmodel1994,PMTgainmodel2017,TAKAHASHI20181-PMT-gainmodel}. While the single p.e.~charge spectrum of the HPK PMT is much narrower than that of the NNVT PMT\,\cite{JUNOPMTgain}. A measured charge spectrum of a single p.e.~(Figure\,\ref{fig:spe}) of NNVT PMT obtained by using a pulsed light source and external trigger (the mean value of the number of impinging photons is around 0.1)\,\cite{PMTgainmodel1994,Luo_2019}. Each simulated event's charge is convoluted through a random sampling process using the single p.e.~spectrum. During the sampling, only events with an amplitude higher than 0.4\,p.e.~in the measured spectrum are considered to remove the non-signal entries. The observed disparity in the charge spectra of simulation between the NNVT and HPK PMTs predominantly arises from the charge smearing applied to the NNVT PMT.

\section{Simulation}
\label{1:sim}
The Geant4 simulation framework \cite{geant4} is employed to simulate the behavior of 20-inch PMTs in the presence of natural radioactivity and cosmic muons traversing the PMT glass bulb. These simulations play a crucial role in enhancing our understanding of the experimental measurements by providing insights into the related physics processes and interactions. Geant4 allows for the accurate modeling of particle interactions and the propagation of radiation through matter, enabling us to simulate the response of the PMTs to various sources of radiation. By comparing the simulation results with the experimental data, we can validate the simulation models and gain valuable insights into the performance of the 20-inch PMTs under different conditions.

\subsection{Thermal Emission}
\label{1:sim:thermal}

The primary source of dark count in PMTs is the thermal electron emission from the PMT photocathode in a dark environment. This emission process, known as thermal emission, involves the spontaneous transfer of internal energy from the thermal reservoir to electrons. The control and understanding of thermal emission are crucial due to its significance and its presence in various applications, including 20-inch PMTs where it serves as the main source of dark count.

\par Thermal electron emissions from the PMT photocathode events are typically independent of each other, and their occurrences are generally consistent with a random distribution. To describe the interval between neighboring thermal electron emissions, an exponential distribution model is often adopted. In this study, we applied the exponential distribution model as a toy Monte Carlo (MC) simulation to numerically characterize the contribution of thermal electrons to the dark count rate in PMTs.

\par By employing this model, we aim to gain further insight into the behavior and characteristics of thermal electron emissions in PMTs, enhancing our understanding of the dark count phenomenon and its underlying mechanisms.

\subsection{Natural Radioactivity}
\label{1:sim:radioactivity}

Radionuclides emit alpha, beta particles, and also gamma radiation. Radioactive materials are widespread in nature, found in soil, underground rock layers, in water, the atmosphere, as well as plants and animals. The primary nuclides of interest include isotopes of uranium (U), thorium (Th), radium-226 (Ra-226), potassium-40 (K-40), polonium-210 (Po-210), lead-210 (Pb-210), tritium (H-3), and others. These radioactive nuclides maintain a dynamic equilibrium of natural radioactivity over an extended period.

\par
\begin{figure*}[!ht]
	\centering
	\subfigure[HPK PMT]{
        \includegraphics[width=0.35\hsize]{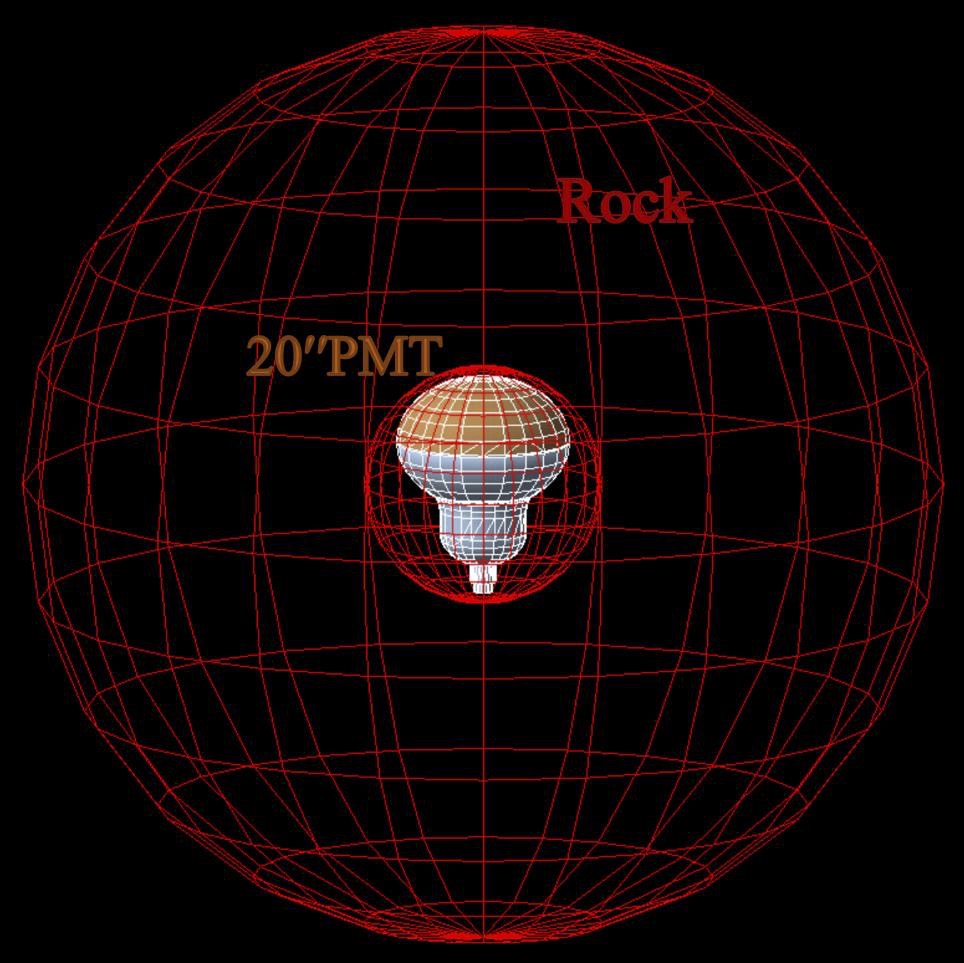}
        %\caption{MCP}
        \label{fig:sim-hpk}
    }
    \quad
	\subfigure[NNVT PMT]{
        \includegraphics[width=0.35\hsize]{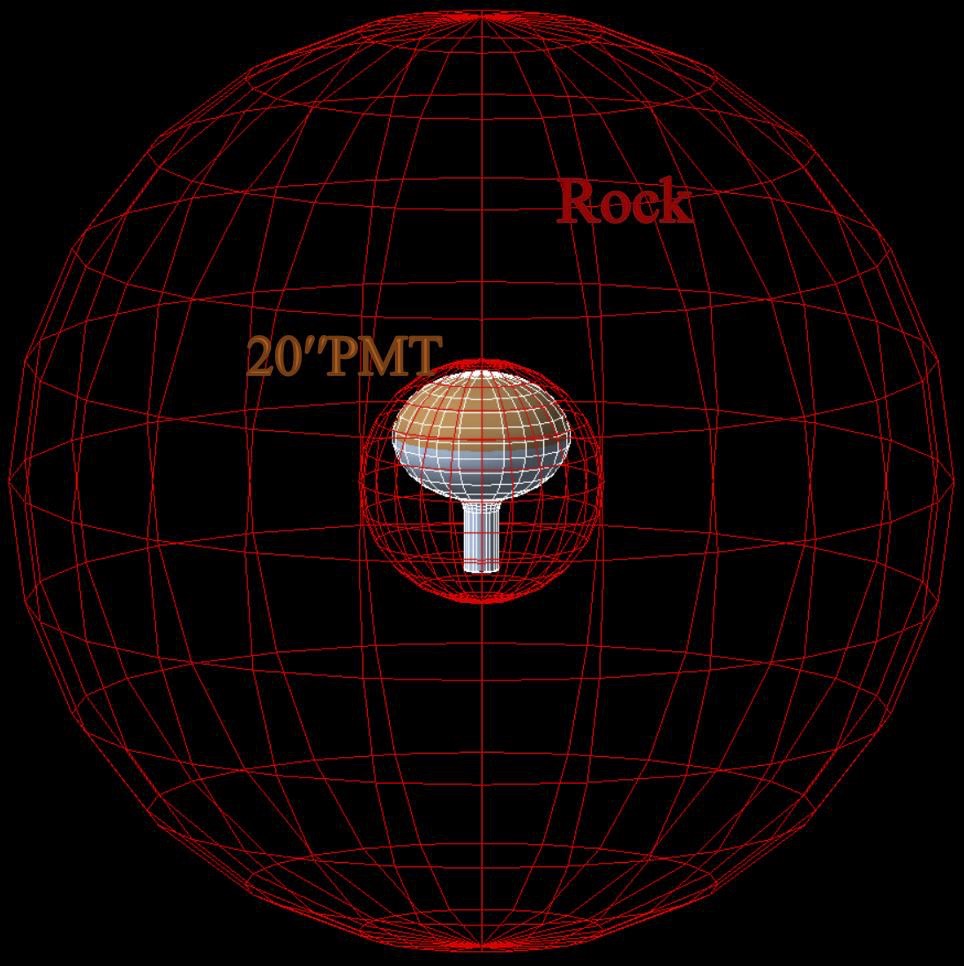}
        \label{fig:sim-nnvt}
    }
    \caption{The simulation of the geometry for the 20-inch PMTs, based on Geant4, was realized for both the (a) HPK and (b) NNVT PMTs.}
    \label{fig:geometry-model}
\end{figure*}

\begin{figure*}[!hbt]
	\centering
	\subfigure[HPK PMT w/o charge smearing]{
        \includegraphics[width=0.4\hsize]{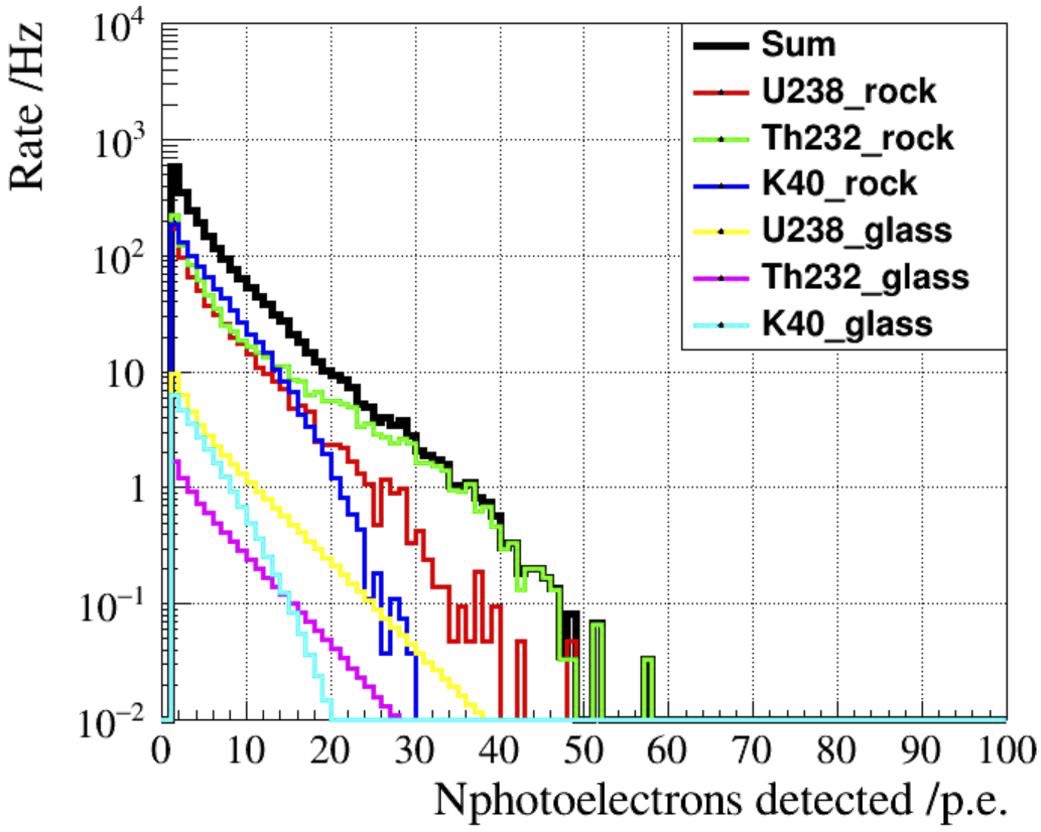}
        %\caption{MCP}
        \label{fig:rad-hpk}
    }
    \quad
	\subfigure[NNVT PMT w/ charge smearing]{
        \includegraphics[width=0.4\hsize]{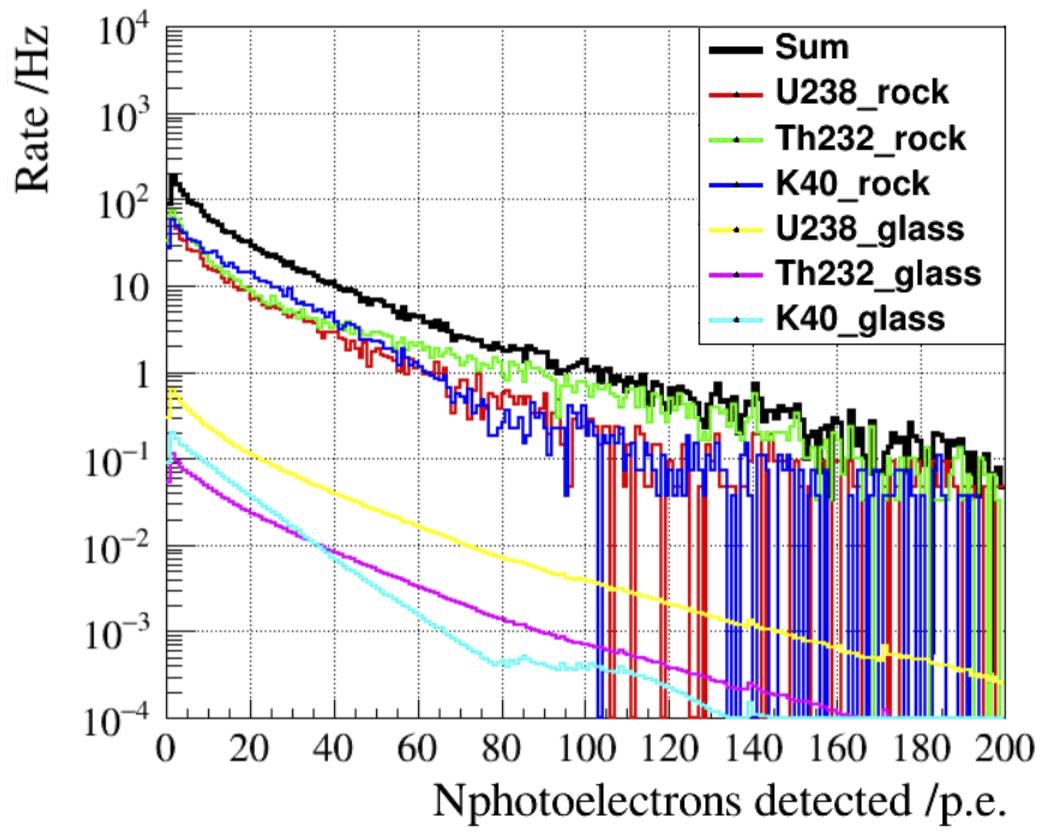}
        \label{fig:rad-nnvt}
    }
    \caption{The responses of 20-inch PMTs to natural radioactivity are depicted in (a) for the HPK PMT and (b) for the NNVT PMT. The difference between the two PMTs arises from the charge smearing effect, which is specific to each PMT's response model. Interestingly, the charge spectra of the NNVT PMT without charge smearing closely resemble those of the HPK PMT as observed in the simulation results. The charge smearing effect of the NNVT PMT is further elaborated upon in section \ref{1:largepulse} to explain the associated processes.}
    \label{fig:rad-sim}
\end{figure*}

 In this study, we consider two sources of natural radioactivity: the PMT glass and the surrounding environment. To simulate this, we employed a simplified geometry where a generator containing $^{238}U/^{232}Th/^{40}K$ isotopes was placed in a uniform thickness (3 mm) of PMT glass, with an additional uniform thickness of 1 m of rock surrounding the PMT (the rock is selected for an equal model only), as depicted in Figure \ref{fig:geometry-model}. The chosen geometry represents an approximation of a realistic configuration, incorporating the concrete material of the experimental room, and enables the generation of uniform natural radioactivity in a 4$\pi$ distribution. The simulation process was validated by considering the PMT glass properties, including Cherenkov radiation and the quantum efficiency (QE) curve of the photocathode. The $^{238}U/^{232}Th/^{40}K$ isotopes were randomly generated, following a 4$\pi$ solid angle distribution on the shell of the rock and PMT glass. The simulation results, shown in Figure \ref{fig:rad-sim}, illustrate the response of the 20-inch PMT to these radioactive sources with the activity and simulated time duration as detailed in Table \ref{tab:rate}. The charge spectra originating from the PMT glass are nearly identical to those from the rocks, with only variations in rates.

\begin{table*}[!htb]
\centering
\caption{The activity of natural radioactivity originating from the surrounding rock and PMT glass was assessed in this study. The simulation duration for glass of HPK and NNVT PMT was determined based on their own radioactivity level\,\cite{ZHANG201867-MCP-glass} to equivalent their simulated samples. }
\label{tab:rate}
\resizebox{0.95\textwidth}{!}{
\begin{tabular}{cc|ccc|ccc|c|c}
\hline
\multicolumn{2}{c|}{\multirow{2}{*}{Natural Radioactivity}}                & \multicolumn{3}{c|}{Rock} & \multicolumn{3}{c|}{Glass} & \multirow{2}{*}{Noise} & \multirow{2}{*}{Muon} \\ \cline{3-8}
\multicolumn{2}{c|}{}  & \multicolumn{1}{c|}{U238}     & \multicolumn{1}{c|}{Th232}  & K40   & \multicolumn{1}{c|}{U238}      & \multicolumn{1}{c|}{Th232}     & K40   &    &    \\ \hline
\multicolumn{1}{c|}{\multirow{2}{*}{Content (Sample measurement)}} & HPK  & \multicolumn{1}{c|}{\multirow{2}{*}{124 Bq/kg}} & \multicolumn{1}{c|}{\multirow{2}{*}{121.5 Bq/kg}} & \multirow{2}{*}{1355 Bq/kg} & \multicolumn{1}{c|}{5 Bq/kg} & \multicolumn{1}{c|}{1.6 Bq/kg} & 11 Bq/kg & /  & /  \\ \cline{2-2} \cline{6-10} 
\multicolumn{1}{c|}{}  & NNVT & \multicolumn{1}{c|}{}  & \multicolumn{1}{c|}{}  &  & \multicolumn{1}{c|}{0.93 Bq/kg} & \multicolumn{1}{c|}{0.3 Bq/kg}  & 0.96 Bq/kg & /  & / \\ \hline
\multicolumn{1}{c|}{\multirow{2}{*}{Rate (Sample measurement)}}    & HPK  & \multicolumn{1}{c|}{600 Hz}  & \multicolumn{1}{c|}{786 Hz}      & 817 Hz   & \multicolumn{1}{c|}{41 Hz}   & \multicolumn{1}{c|}{8.2 Hz}    & 26 Hz   & 5.6 kHz                & 44 Hz               \\ \cline{2-10} 
\multicolumn{1}{c|}{}                                               & NNVT & \multicolumn{1}{c|}{592 Hz}                  & \multicolumn{1}{c|}{775 Hz}                    & 785 Hz                   & \multicolumn{1}{c|}{7.8 Hz}    & \multicolumn{1}{c|}{1.5 Hz}    & 2.3 Hz    & 8.3 kHz                & 39 Hz               \\ \hline
\multicolumn{1}{c|}{\multirow{2}{*}{Statistical   Magnitude (s)}}   & HPK  & \multicolumn{1}{c|}{\multirow{2}{*}{21}}    & \multicolumn{1}{c|}{\multirow{2}{*}{30}}      & \multirow{2}{*}{27}     & \multicolumn{1}{c|}{3.2 $\times$ 10$^5$}    & \multicolumn{1}{c|}{1.4 $\times$ 10$^6$}   & 2.1 $\times$ 10$^6$   & /  & /  \\ \cline{2-2} \cline{6-10} 
\multicolumn{1}{c|}{}   & NNVT & \multicolumn{1}{c|}{}  & \multicolumn{1}{c|}{}   &   & \multicolumn{1}{c|}{1.8 $\times$10$^6$}   & \multicolumn{1}{c|}{7.6 $\times$10$^6$}   & 2.4 $\times$10$^7$  & /  & /   \\ \hline
\end{tabular}
}
\end{table*}

It is worth to note that the charge spectra of NNVT PMT without charge smearing (considering the real charge response of the PMT) are almost the same as that from HPK PMT from simulation, while it is no obviously difference before and after the charge smearing of the HPK PMT. The charge smearing of NNVT PMT is further discussed in section\,\ref{1:largepulse} for the related process. The $^{238}U/^{232}Th/^{40}K$ outside of the 1\,m rock shell has significantly low contribution to the results and can be ignored.

\subsection{Cosmic Muon}
\label{1:sim:muon}

It is well-known that when an incoming muon penetrates through the PMT glass, it generates a pulse through the phenomenon of Cherenkov radiation. This phenomenon has been extensively studied in previous works \cite{PMTmuon2007,BAYAT20141}. The large pulses observed in the 20-inch PMT dark count can be attributed to the photons generated by muons passing through the PMT glass via Cherenkov radiation. This occurs when the muon's speed exceeds the phase velocity of light in the glass. Comparing the experimental data with theoretical predictions, it becomes evident that there is a significant increase in the pulse rate, exceeding 10 Hz, when the threshold is set above approximately 3 p.e.~This rate can even exceed 10 p.e.~for both HPK and NNVT PMTs \cite{zhang2022study}.

\par To further investigate this phenomenon, we employed geometry models of the 20-inch NNVT PMT and HPK PMT. Muons were generated randomly according to a muon flux at sea level\,\cite{Guan2015APO-muonflux,PhysRevD.58.054001-muonflux} from a plane of 10$\times$10 m$^{2}$, located above the PMT. We performed simulations to analyze the PMT response to muons passing through the PMT glass \cite{zhang2022study}. The resulting charge curves were then used for subsequent analysis. Our results confirm that the large pulses observed in the 20-inch PMT dark count primarily originate from the photons generated by muons that penetrate the PMT glass through Cherenkov radiation, regardless of the PMT type (HPK or NNVT). For a more comprehensive understanding, please refer to \cite{zhang2022study}.

\section{Comparison with Measurement}
\label{1:comp}

By employing Geant4 and toy MC simulations, we investigated the behavior of a PMT in a dark environment, considering the thermal electron emission, natural radioactivity, and cosmic muons. These simulations provided valuable insights into the structure of the PMT dark count rate (DCR) and charge spectrum. However, to gain a comprehensive understanding and accurately determine the contribution of each component, further comparisons with experimental measurements are required.

%\par The preliminary findings from the simulations have shed light on the overall characteristics of the PMT DCR charge spectrum. However, a more detailed analysis is necessary to explore the individual contributions of thermal electron emission, natural radioactivity, and cosmic muons when comparing the simulation results with experimental data. This comparative analysis will allow us to disentangle the distinct signatures and quantify the impact of each component on the PMT DCR.

\par Through this comprehensive comparison between simulation and measurement, we aim to refine our understanding of the relative importance of thermal electron emission, natural radioactivity, and cosmic muons in shaping the PMT DCR charge spectrum. This knowledge will enhance our ability to accurately model and predict the behavior of PMTs in dark conditions.

\subsection{Fitting Model}
\label{1:comp:fit}

\begin{figure*}[!htb]
	\centering
	\subfigure[HPK PMT]{
        \includegraphics[width=0.45\hsize]{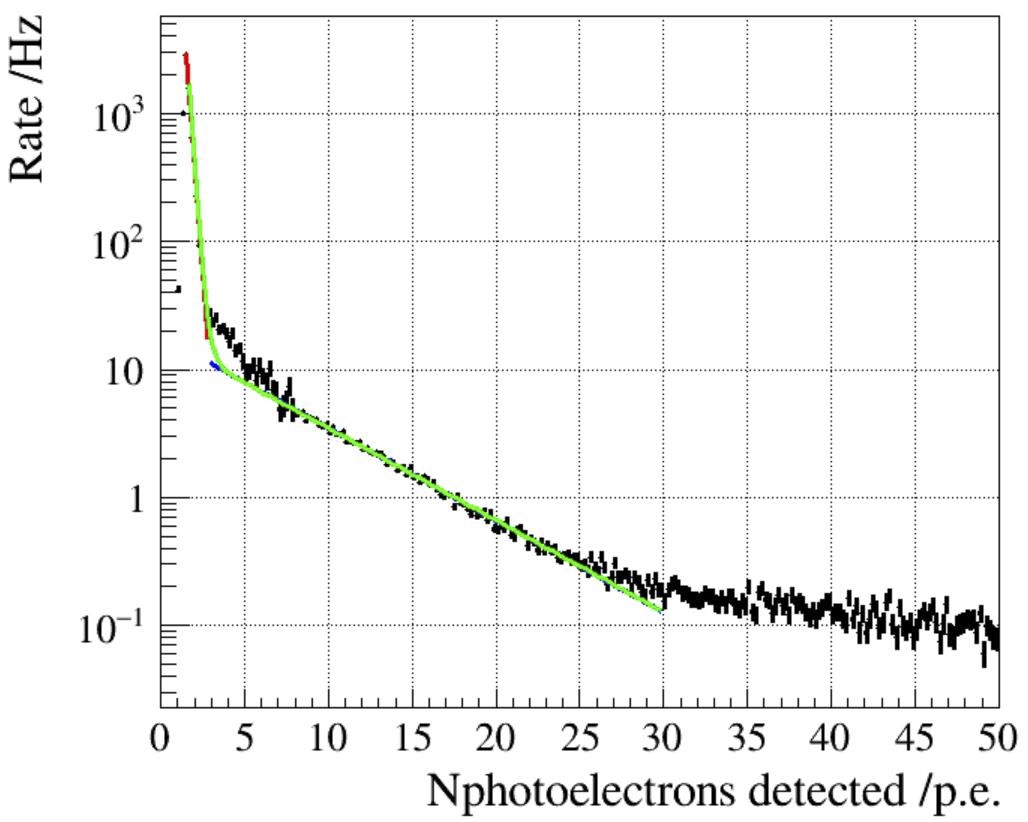}
        %\caption{MCP}
        \label{fig:Thermal emission}
    }
    \quad
	\subfigure[NNVT PMT]{
        \includegraphics[width=0.45\hsize]{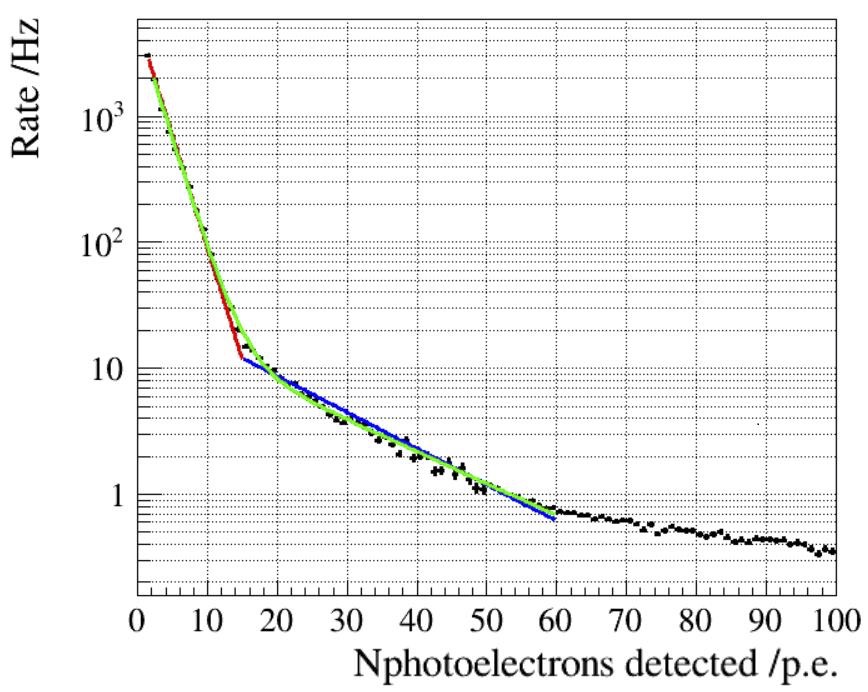}
        \label{fig:Thermal emission-nnvt}
    }
    \caption{The fitting results of the thermal electron emission in the measured charge spectrum of the 20-inch PMTs are shown in (a) for the HPK PMT and (b) for the NNVT PMT. Different fitting ranges are suggested for different charge response feature (also related to the applied gain model as stated) of the two kind of PMTs, which affects the rate distribution even they are contributed by the same source. For example, the thermal noise is only extends to around 3 p.e.\,for the HPK PMT, but it is around 15 p.e.\,for the NNVT PMT. There has a similar effect on the contribution of the natural radioactivity (3-30 p.e.\,for the HPK PMT, but 15-60 p.e.\,for the NNVT PMT). In the case of the HPK PMT, the fitting range spans from 1 p.e.~to 30 p.e., represented by the green line. For the NNVT PMT, the fitting range extends from 1 p.e.~to 60 p.e. }
    \label{fig:sim-model}
\end{figure*}

In Section \ref{1:sim:thermal}, we discussed the fitting of the thermal electron emission in the charge range of 1\,p.e.~to 3\,p.e.~for the HPK PMT (and extended to around 10\,p.e.~for the NNVT PMT considering their charge smearing, as shown in Figure \ref{fig:Thermal emission-nnvt}) for the measured data. The fitting process aims to accurately capture the characteristics of the thermal electron emission mechanism.

\par To ensure accurate fitting results and avoid the confounding effects of natural radioactivity and thermal noise, a hypothetical exponential distribution of natural radioactivity is introduced in the transition section after the thermal noise. This approach allows for a smoother and more precise fitting of the thermal noise by only one exponential function. The fitting function of the thermal noise can be expressed as follows:
\begin{equation}\label{con:Thermal_emission}
    f\left( x\right) =A_{1} \times e^{b_{1}x}+A_{2} \times e^{b_{2}x}
     \end{equation}
\par Where $A_{1}$ and $b_{1}$ are used for the contribution of the thermal noise, and $A_{2}$ and $b_{2}$ are used for the contribution of the natural radioactivity. By adopting this fitting method, we obtain a realistic and reliable estimation of the thermal noise, which is essential for understanding and modeling the thermal electron emission in 20-inch PMTs. The fitting results provide valuable insights into the behavior of the PMT in the low charge range and contribute to our overall understanding of the PMT characteristics.

\par The fitting results for the thermal emission of the 20-inch HPK PMT indicate that the distribution can be described by $A_{1}$=$1.365 \times 10^{6}$, and $b_{1}$=-4.08, which dominates in the low part of the fitting range. This distribution corresponds to a dark count rate of approximately 43.5\,kHz when integrated over the distribution and considering a threshold of 0.5\,p.e., aligning well with the measurement data. On the other hand, the green lines in Figure \ref{fig:sim-model} represent the estimated radioactivity contribution, modeled by $A_{2}$=$1.81 \times 10^{1}$, and $b_{2}$=-0.165 for the 20-inch HPK PMT, resulting in a rate of approximately 101.0\,Hz. The colored lines in the figure represent the results of separate fits for the thermal noise and estimated radioactivity components.

\par For the 20-inch NNVT PMT, the fitted constants are $A_{1}$=$5.20 \times 10^{3}$ and $b_{1}$=-0.416, $A_{2}$=$2.25 \times 10^{1}$ and $b_{2}$=-0.0584, corresponding to a dark count rate of 10.2\,kHz for thermal noise and 374.2\,Hz for estimated radioactivity, respectively, when considering a threshold of 0.5\,p.e. 

\par These fitting results represent the behavior of the PMTs in terms of their thermal noise and estimated radioactivity contributions, and suggest distinct characteristics in their dark count behavior with respect to charge levels, where the different charge response feature (the 20-inch NNVT PMT has a longer tail than the HPK PMT as shown in Figure\,\ref{fig:spe}) between the two kinds of PMTs dominant the left part of the rate vs.\,charge plot. It is also the reason for different fitting range of the thermal noise between the 20-inch NNVT PMT and 20-inch HPK PMT as stated.

The initial values for the fitting process were obtained from the simulated natural radioactivity of $^{238}U/^{232}Th/^{40}K$ in the rock and PMT glass, as shown in Figure \ref{fig:rad-sim}. The TMinuit software package was used to perform the minimization procedure, resulting in the final fitting result. The complete contribution of natural radioactivity consists of six components originating from the glass and surrounding rock, which are superimposed as shown in Eq.\,\ref{con:NaturalRad}.
In this equation, $x_j$ represents the contribution of $N_i^{NaturalRad}$ within the light intensity range of [a,b] p.e.~The fitting of $x_j$ was carried out using Eq.\,\ref{con:chi}, which takes into account the contributions from thermal electron emission and muons. The parameters $x_{Muon}$ and $x_{Noise}$ correspond to the contributions of $N_i^{Muon}$ and $N_i^{Noise}$ within the light intensity range of [a,b] p.e., respectively.

\begin{figure*}[!hbt]
    \centering
    \begin{align}
    \label{con:NaturalRad}
    N_{i}^{NaturalRad}&=\sum ^{6}_{j=1}x_{j}N_{j} \\
    \label{con:chi}
    \chi^{2}&=\sum ^{b}_{i=a}\dfrac{\left( N_{i}^{NaturalRad}+x_{Muon} \times N_{i}^{Muon}+x_{Noise} \times N_{i}^{Noise}-N_{i}^{Test}\right) ^{2}}{N_i^{test}} 
    \end{align}
\end{figure*}

Based on the fitting results and the covariance matrix of parameters, it was observed that the natural radioactivity of the glass and rock, represented by $x_j$, exhibited a strong correlation with each other.

To further refine the analysis, the natural radioactivity contribution from the glass was fixed, considering its much smaller contribution compared to the surrounding rock. This was performed using two strategies: (1) fixing only the three parameters related to natural radioactivity ($^{238}U/^{232}Th/^{40}K$) in the rock, while fixing the thermal electron emission, glass, and muon parameters (mode \#1), and (2) incorporating thermal electron emission and muon parameters to have a total of five parameters (mode \#2).

In order to obtain accurate results, it is necessary to consider all the components that contribute to the full spectrum, including noise, natural radioactivity, and muon. However, it is also important to take into account the distortion around the amplitude threshold in the data.

\par For the HPK PMT, various fitting ranges were explored within the ranges of [3,80]\,p.e.~and [5,80]\,p.e.~(see table~\ref{tab:HPK-table}). In the case of the NNVT PMT, wider fitting ranges were tested (see table~\ref{tab:MCP-table}).

For the HPK PMT, both fitting modes \#1 and \#2 were examined, and they yielded similar fitting quality. However, there were slight differences in the values and uncertainties of each decay chain of $^{238}U/^{232}Th/^{40}K$, as indicated in table~\ref{tab:HPK-table}. The $^{238}U/^{232}Th/^{40}K$ in the PMT glass was kept fixed during the fitting within the range of [5,80]\,p.e., as shown in Figure\,\ref{fig:fit5-80hpk}. Mode \#2 included five parameters: $^{238}U/^{232}Th/^{40}K$ in the rock, noise, and cosmic muon, which were compared with the experimental data.
\begin{figure*}[!htb]
    \centering
    \subfigure[HPK PMT fitted curves]{
    \includegraphics[width=0.45\hsize]{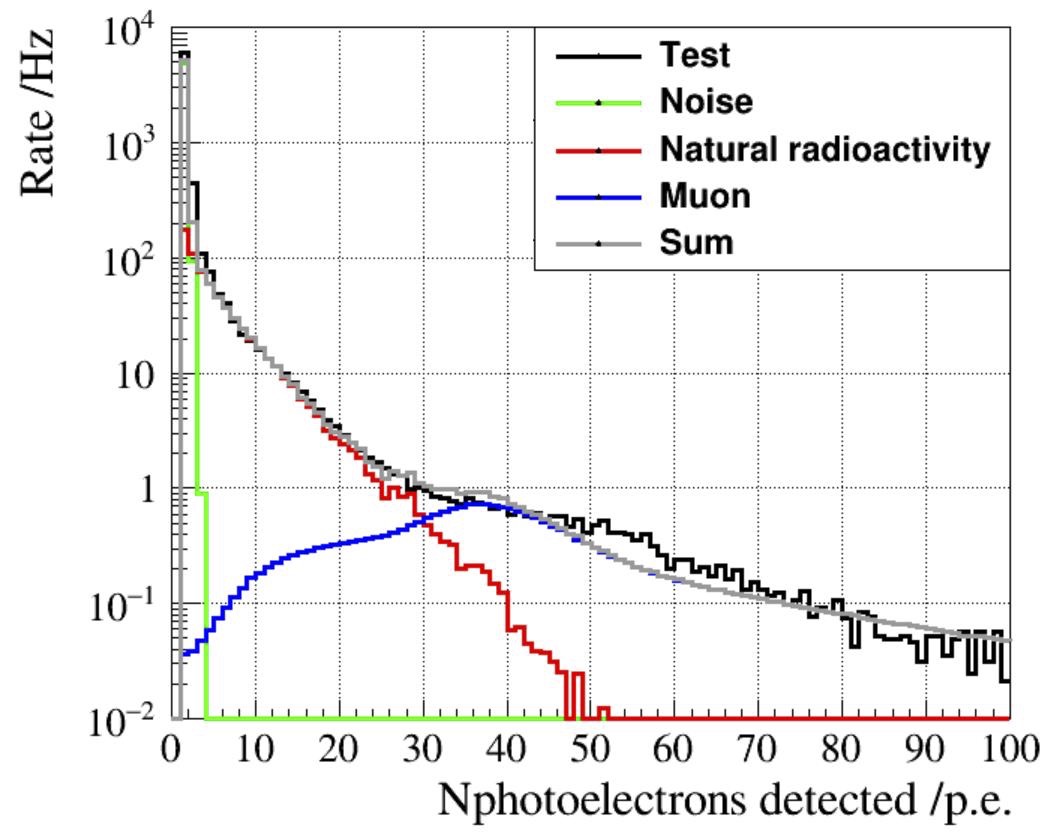}
    %\caption{MCP}
    \label{fig:fit5-80hpk}
    }
    \quad
    \subfigure[NNVT PMT fitted curves]{
    \includegraphics[width=0.45\hsize]{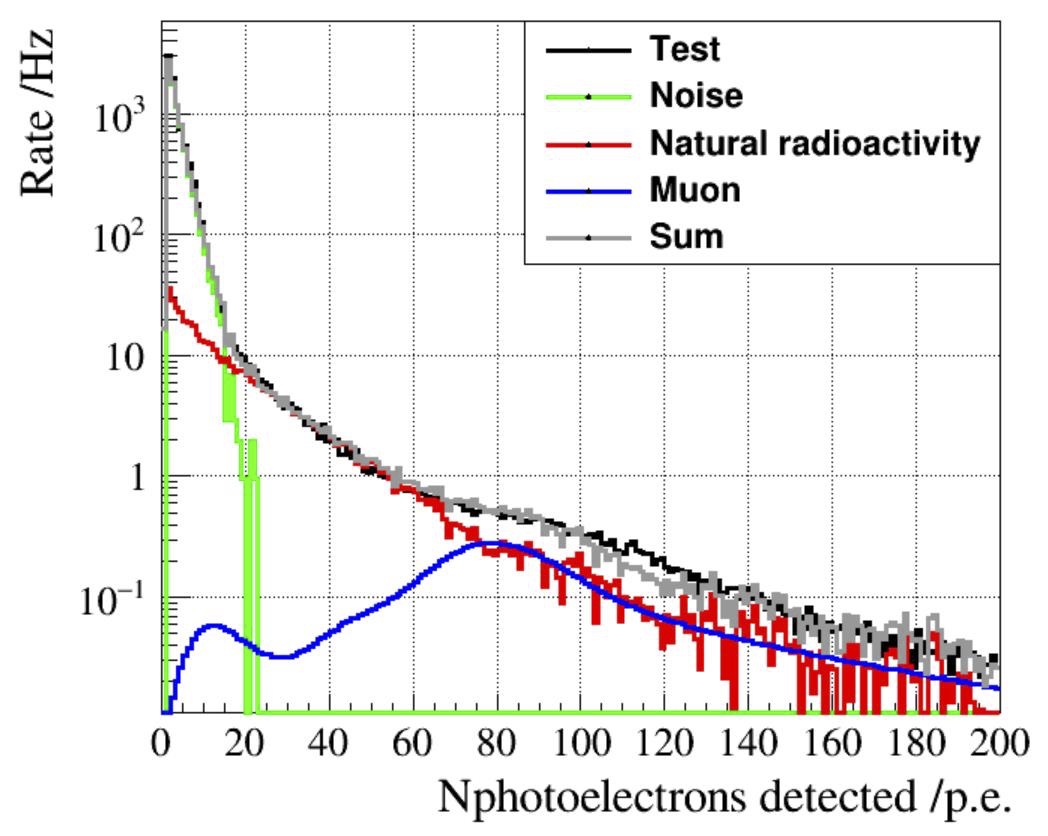}
    \label{fig:fit3-140-nnvt}
    }
    \caption{The fitted curves of (a) HPK and (b) NNVT PMTs encompass various components, including noise, muon, and natural radioactivity. Note: the different positions of the muon peaks on x-axes is from different charge response features between HPK and NNVT PMTs, and the applied gain model used to convert charge to p.e.\,as stated.}
    \label{fig:sim-model-2}
\end{figure*}

\par For the NNVT PMT, the fitting results of mode \#1 showed insignificant changes for different fitting ranges, while mode \#2 was further adjusted using several fitting ranges. Although slight differences were observed in the fitting parameters for different ranges, the overall fitting quality remained relatively unchanged, as listed in table~\ref{tab:MCP-table}. In mode \#2, the $^{238}U/^{232}Th/^{40}K$ in the PMT glass was fixed during the fitting within the range of [3,140]\,p.e., as depicted in Figure\,\ref{fig:fit3-140-nnvt}. Five parameters were considered, including $^{238}U/^{232}Th/^{40}K$ in the rock, noise, and cosmic muon, which were compared with the available data.

\begin{table*}[!htb]
\centering
\caption{The fitting results of the HPK PMT are obtained by applying multiple fitting ranges to cover different contributions and verify the fitting quality. By utilizing various fitting ranges, we aim to account for the diverse sources of signal and assess the reliability of the fitting process. Note: the fitting quality and parameter uncertainty is heavily related to the global fitting range (especially the lower limit), and it is the reason why it is tried with different configurations.}
\label{tab:HPK-table}
\resizebox{0.95\textwidth}{!}{
\begin{tabular}{cc|c|c|c|c|c|c|c|c}
\hline
\multicolumn{2}{c|}{Rate   (Hz)}                                         & Rock-U238     & Rock-Th232    & Rock-K40      & Glass-U238 & Glass-Th232 & Glass-K40 & Noise          & Muon       \\ \hline
\multicolumn{1}{c|}{ Only free rock }   & [3,80]p.e. & 97±38   & 199±36  & 361±34  & /         & /           & /         & /              & /          \\ \cline{2-10} 
\multicolumn{1}{c|}{}                                       & [5,80]p.e. & 176 ± 45 & 164 ± 42 & 261 ± 40  & /         & /           & /         & /              & /          \\ \hline
\multicolumn{1}{c|}{Rock/Noise/Muon} & [3,80]p.e. & 138 ± 430 & 167 ± 197 & 317 ± 242 & /         & /           & /         & 150k ± 60k & 27 ± 8.7 \\ \cline{2-10} 
\multicolumn{1}{c|}{}                                       & [5,80]p.e. & 220 ± 45 & 134 ± 42 & 247 ± 40  & /         & /           & /         & 5.0k ± 0.8k & 28 ± 7.1 \\ \hline
\end{tabular}
}
\end{table*}

\begin{table*}[!htb]
\centering
\caption{The fitting results of the NNVT PMT in mode \#2 are obtained by applying multiple fitting ranges to cover different contributions and verify the fitting quality. By testing various fitting ranges, we aim to encompass the different sources of signal and ensure the accuracy of the fitting process.} 
\label{tab:MCP-table}
\resizebox{0.95\textwidth}{!}{

\begin{tabular}{cc|c|c|c|c|c|c|c|c}
\hline
\multicolumn{2}{c|}{Rate   (Hz)}    & Rock-U238     & Rock-Th232   & Rock-K40      & Glass-238 & Glass-Th232 & Glass-K40 & Noise          & Muon        \\ \hline
\multicolumn{1}{l|}{} & [3,120]p.e. & (1.6 ± 2.3)$\times 10^2$ & (0.5 ± 1.8)$\times 10^2$ & (2.3 ± 1.8)$\times 10^2$ & / & / & / & (7.7 ± 0.1)$\times 10^3$ & (0.15 ± 0.11)$\times 10^2$ \\ \cline{2-10} 
\multicolumn{1}{l|}{} & [5,120]p.e. & (1.1 ± 2.3)$\times 10^2$ & (0.4 ± 2.0)$\times 10^2$ & (2.6 ± 1.8)$\times 10^2$ & / & / & / & (8.4 ± 0.2)$\times 10^3$ & (0.19 ± 0.14)$\times 10^2$ \\ \cline{2-10} 
\multicolumn{1}{c|}{Fit Rock} & [3,140]p.e. & (1.4 ± 2.2)$\times$10$^2$ & (0.63 ± 1.6)$\times 10^2$ & (2.4 ± 1.7)$\times 10^2$ & / & / & / &  (7.7 ± 0.1)$\times 10^3$ & (0.15 ± 0.10)$\times 10^2$ \\ \cline{2-10}

\multicolumn{1}{c|}{/Noise} & [5,140]p.e. & (0.0 ± 3.7)$\times 10^4$ & (0.51 ± 1.8)$\times 10^2$ & (3.2 ± 2.5)$\times$10$^2$ & / & / & / & (8.6 ± 0.3)$\times$10$^3$ & (0.23 ± 0.16)$\times 10^2$ \\ \cline{2-10} 
\multicolumn{1}{c|}{/Muon} & [3,100]p.e. & (0.9 ± 2.5)$\times 10^2$ & (0.3 ± 3.0)$\times 10^2$ & (2.9 ± 1.9)$\times$ 10$^2$ & /   & /    & /         &  (8.0 ± 0.1)$\times$ 10$^3$ & (0.17 ± 0.13)$\times$ 10$^2$ \\ \cline{2-10} 
\multicolumn{1}{l|}{} & [3,80]p.e.  & (1.7 ± 2.9)$\times 10^2$ & (0.0 ± 5.8)$\times 10^3$ & (2.7 ± 2.1)$\times 10^2$ & /  & /  & /   & (7.8 ± 0.1)$\times 10^3$ & (0.13 ± 0.16)$\times 10^2$ \\ \cline{2-10} 
\multicolumn{1}{l|}{} & [2,140]p.e. & (0.6 ± 2.2)$\times 10^2$ & (0.3 ± 2.2)$\times 10^2$ & (3.2 ± 1.8)$\times 10^2$ & / & /   & /  & (8.2 ± 0.1)$\times 10^3$ & (0.21 ± 0.15)$\times 10^2$ \\ \hline
\end{tabular}

}
\end{table*}

\iffalse
\begin{figure*}[!htb]
    \centering
    \subfigure[HPK PMT fitted curves]{
    \includegraphics[width=0.35\hsize]{figure/fit5-80hpk.jpg}
    %\caption{MCP}
    \label{fig:fit5-80hpk}
    }
    \quad
    \subfigure[Fitted natural activity]{
    \includegraphics[width=0.35\hsize]{figure/fit5-80-hpk.jpg}
    \label{fig:fit5-80-hpk}
    }
    \caption{HPK PMT fitted curves, contains noise, muon, natural radioactivity (individual component) and test contrast. (a) Comparison of simulated muon, thermal electron emission and natural radioactivity with test results for HPK PMT; (b) Simulated natural radioactive components in HPK PMT. \textcolor{green}{((b) already showed in Fig3? repeated? combine with fig 7?)}}
    \label{fig:sim-model-2}
\end{figure*}

\begin{figure}[!htb]
    \centering
    \includegraphics[width=0.76\hsize]{figure/fit3-140-nnvt.jpg}
    \caption{The DCR charge spectrum fitted curves of NNVT PMT, contains noise, muon, natural radioactivity and test contrast.\textcolor{green}{(difference to fig 6 and 7?)}}
    \label{fig:fit3-140-nnvt}
\end{figure}
\fi

\subsection{Comparing with Measured Radioactivity}
\label{1:comp:expectation}

Considering the fitted errors obtained from all the fitting results, it is worth noting that the NNVT PMT exhibits larger errors compared to the HPK PMT, which is heavily related to the fitting range especially the lower limit. Due to this discrepancy, it is not appropriate to express the overall fitting result simply as an arithmetic average. Instead, a weighted average is employed to account for the variations in the fitting results. Each fitting result is denoted as $L_i$, with an associated error of $m_i$, and the weight is calculated as $p_i=\frac{1}{m_i}$. Consequently, the weighted average $\overline{x}_i$ and the corresponding error $m_x$ can be defined as follows:
\begin{equation}
\begin{split}
  %  \begin{align}
    \label{con:mean}
    \overline{x}_i&=\dfrac{\sum_{i=1}^{n}p_{i}l_{i}}{\sum^{n}_{i=1}p_{i}} \\%\label{con:error}
    m_{x}&=\sqrt{\dfrac{1}{\sum ^{n}_{i=1}p_{i}}}
 %   \end{align}
\end{split}
\end{equation}

\begin{figure*}[!ht]
    \centering
    \subfigure[HPK PMT]{
    \includegraphics[width=0.46\hsize]{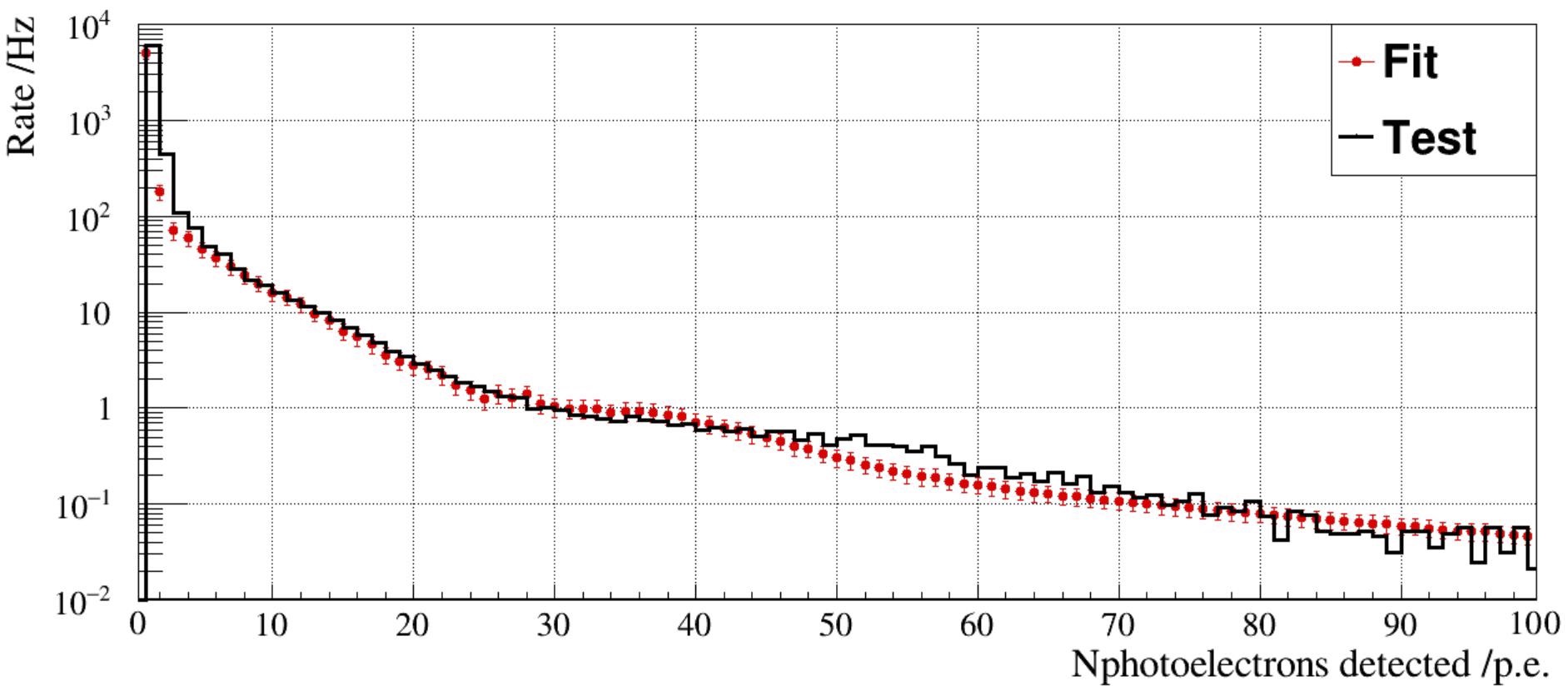}
    %\caption{MCP}
    \label{fig:hpk-sum}
    }
    \quad
    \subfigure[NNVT PMT]{
    \includegraphics[width=0.46\hsize]{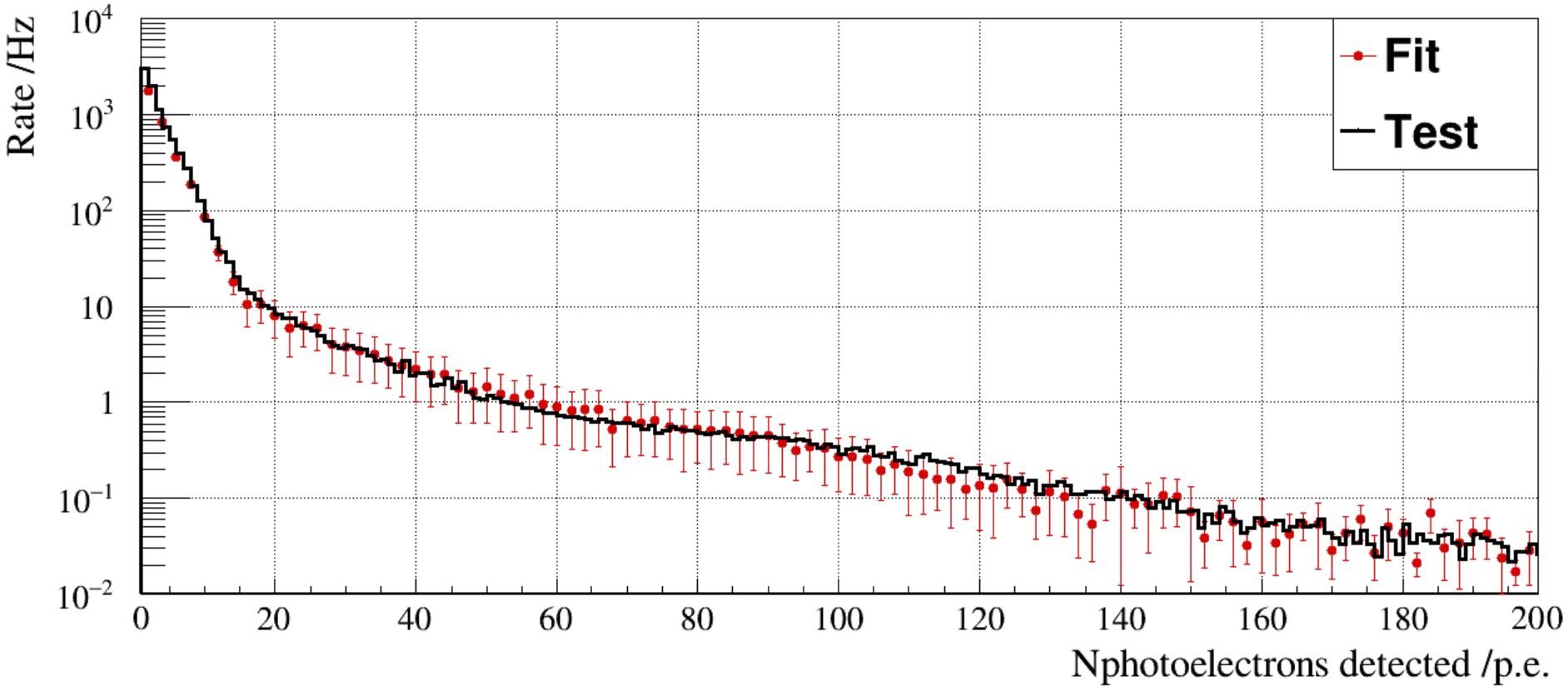}
    \label{fig:nnvt-sum}
    }
    \caption{The expected DCR charge spectrum with the averaged fitted parameters for (a) HPK and (b) NNVT PMTs, which take into account noise, muon and, natural radioactivity. The error of each parameter is considered into the uncertainty of the expected spectra. Note: the first kink (around 3 p.e.\,for HPk PMT and 15 p.e.\,for NNVT PMT) is mainly from the contribution of natural radioactivity; the second kink (around 25 p.e.\,for HPk PMT and 70 p.e.\,for NNVT PMT) is mainly from the contribution of the muons.}
    \label{fig:sim-sum}
\end{figure*}

The compositions of $^{238}U/^{232}Th/^{40}K$ in the rock, as well as the rates of each component obtained from the weighted average strategy, are presented in table~\ref{tab:sum1} in radioactivity and rates, and comparison of sample measurements is shown in table~\ref{tab:rate}. A comprehensive comparison between the measurement and the expected values, incorporating the averaged compositions, can be observed in Figure\,\ref{fig:sim-sum}. The plots also take into account the errors associated with the parameters.

\par The NNVT and HPK PMTs demonstrate a similar trend following Eq.\,\ref{con:mean} within their relatively large uncertainties in the results. However, when comparing the fitted values with the initial values, a notable reduction is observed in the $^{238}U/^{232}Th/^{40}K$ composition of the rock.

\par It should be noted that the difference between the laboratory testing environment and the simulation setting might contribute to these variations. The natural radioactivity of the surrounding material, as obtained from the fitting process, aligns more closely with the concrete model\,\cite{armengaud2017performance} rather than the rock itself. Additionally, the simulation configuration assumes an ideal scenario with 4$\pi$ coverage, which could further contribute to the observed differences.

\begin{table*}[!htb]
\centering
\caption{Summary in radioactivity of measurement fitting: HPK and NNVT PMT Comparison. The next section needs to predict the 20-inch PMT spectrum of natural radioactivity in JUNO. The simulated rate of muon is 100 Hz/m$^2$.}
\label{tab:sum1}
\resizebox{0.95\textwidth}{!}{
\begin{tabular}{cc|ccc|ccc|c|c}
\hline
\multicolumn{2}{c|}{\multirow{2}{*}{Fit Rock}}                 & \multicolumn{3}{c|}{Rock}     & \multicolumn{3}{c|}{Glass}     & \multirow{2}{*}{Noise} & \multirow{2}{*}{Muon} \\ \cline{3-8}
\multicolumn{2}{c|}{/Noise/Muon}      & \multicolumn{1}{c|}{U238}         & \multicolumn{1}{c|}{Th232}  & K40     & \multicolumn{1}{c|}{U238}   & \multicolumn{1}{c|}{Th232}  & K40    &   &    \\ \hline

\multicolumn{1}{c|}{\multirow{2}{*}{Content }} & HPK     & \multicolumn{1}{c|}{(0.43 ± 0.09)$\times 10^2$}   & \multicolumn{1}{c|}{(0.21±0.06)$\times 10^2$}   & (4.1±0.7)$\times 10^2$  & \multicolumn{1}{c|}{/}      & \multicolumn{1}{c|}{/}      & /      & /                      & /                     \\ \cline{2-10} 
\multicolumn{1}{c|}{(Bq/kg)}                                             & NNVT       & \multicolumn{1}{c|}{(0.25±0.20)$\times 10^2$}  & \multicolumn{1}{c|}{7±13}   & (4.7 ± 1.2)$\times 10^2$ & \multicolumn{1}{c|}{/}      & \multicolumn{1}{c|}{/}      & /      & /                      & /                     \\ \hline
\multicolumn{1}{c|}{\multirow{4}{*}{Rate (Hz)}}    & \multirow{2}{*}{HPK}  & \multicolumn{1}{c|}{(2.2±0.5)$\times 10^2$}                & \multicolumn{1}{c|}{(1.4±0.4)$\times 10^2$}                & (2.5±0.4)$\times 10^2$                  & \multicolumn{1}{c|}{/}      & \multicolumn{1}{c|}{/}      & /      & (5.1±0.8)$\times 10^3$            & (0.27±0.06)$\times 10^2$            \\   
\multicolumn{1}{c|}{}     &       & \multicolumn{1}{c|}{(3.8\%)}           & \multicolumn{1}{c|}{(2.4\%)}    & (4.4\%)        & \multicolumn{1}{c|}{} & \multicolumn{1}{c|}{} &  & (89\%)    & (0.48\%)    \\ \cline{2-10} 
\multicolumn{1}{c|}{}    & \multirow{2}{*}{NNVT} & \multicolumn{1}{c|}{(1.2±1.0)$\times 10^2$}        & \multicolumn{1}{c|}{(0.48±0.80)$\times 10^2$}     & (2.7±0.7)$\times 10^2$     & \multicolumn{1}{c|}{/}      & \multicolumn{1}{c|}{/}      & /      & (8.0±0.1)$\times 10^3$            & (0.17±0.05)$\times 10^2$            \\  
\multicolumn{1}{c|}{}      &      & \multicolumn{1}{c|}{(1.40\%)}          & \multicolumn{1}{c|}{(0.57\%)}       & (3.2\%)                        & \multicolumn{1}{c|}{} & \multicolumn{1}{c|}{} &  & (94\%)                & (0.20\%)       \\ \hline
\end{tabular}
}
\end{table*}

\section{Underground Prediction}
\label{1:exp} 
The JUNO experiment is expected to encounter several primary background sources, including cosmic muons, fast neutrons, cosmogenic isotopes generated by muon spallation in the liquid scintillator, and natural radioactivity. The strict control of natural radioactivity aims to reduce the accidental count rate in the JUNO detector and minimize low-energy events. Natural radioactivity arises from various materials in the environment. The main source of natural radioactivity detected by PMTs in JUNO is the $^{238}U/^{232}Th/^{40}K$ content present in both JUNO's surrounding rocks (as shown in table~\ref{tab:juno-like}) and the glass components of the PMTs\,\cite{abusleme2021juno}.

\par By applying mode \#2, it is possible to provide preliminary predictions of the 20-inch HPK and NNVT PMT dark noise spectra in the underground environment of JUNO, as depicted in Figure\,\ref{fig:juno-like}. These predictions are based on the assumption of pre-water filling, and it is anticipated that the contribution from radioactivity in the surrounding rocks will be further reduced after filling the detector with pure water. The presence of a thick rock layer underground significantly diminishes the contribution of muons to the charge spectrum compared to muons at the ground level.

At a depth of 700 m, the underground environment effectively suppresses the muon rate, ensuring that no muon contributions are observed in the actual PMT spectrum. The differences between HPK and NNVT PMTs have been taken into account in section\,\ref{1:comp:fit}. Consequently, it is possible to provide preliminary predictions of the 20-inch PMT spectra influenced by natural radioactivity within a specific experimental environment.

\begin{table}[!ht]
\centering
\caption{Natural radioactivity of underground rocks in the JUNO site \cite{junocollaboration2023juno}.}
\label{tab:juno-like}
\begin{tabular}{c|c|c|c}
\hline
Isotope & Rock-U238 & Rock-Th232 & Rock-K40 \\ \hline
Activity (Bq/kg)  & 110±10 & 105±10 & 1340±50 \\ \hline
\end{tabular}
\end{table}

\begin{figure*}[!ht]
	\centering
	\subfigure[HPK PMT]{
        \includegraphics[width=0.45\hsize]{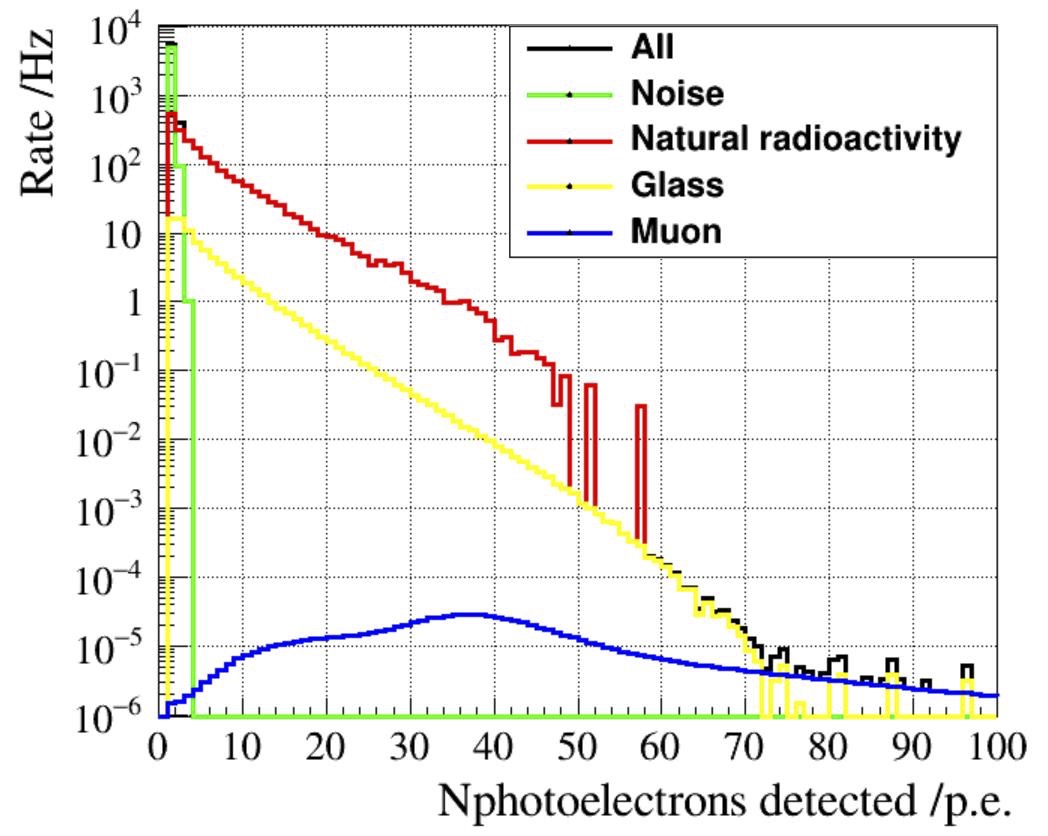}
        \label{fig:juno-hpk}
    }
    \quad 
	\subfigure[NNVT PMT]{
        \includegraphics[width=0.45\hsize]{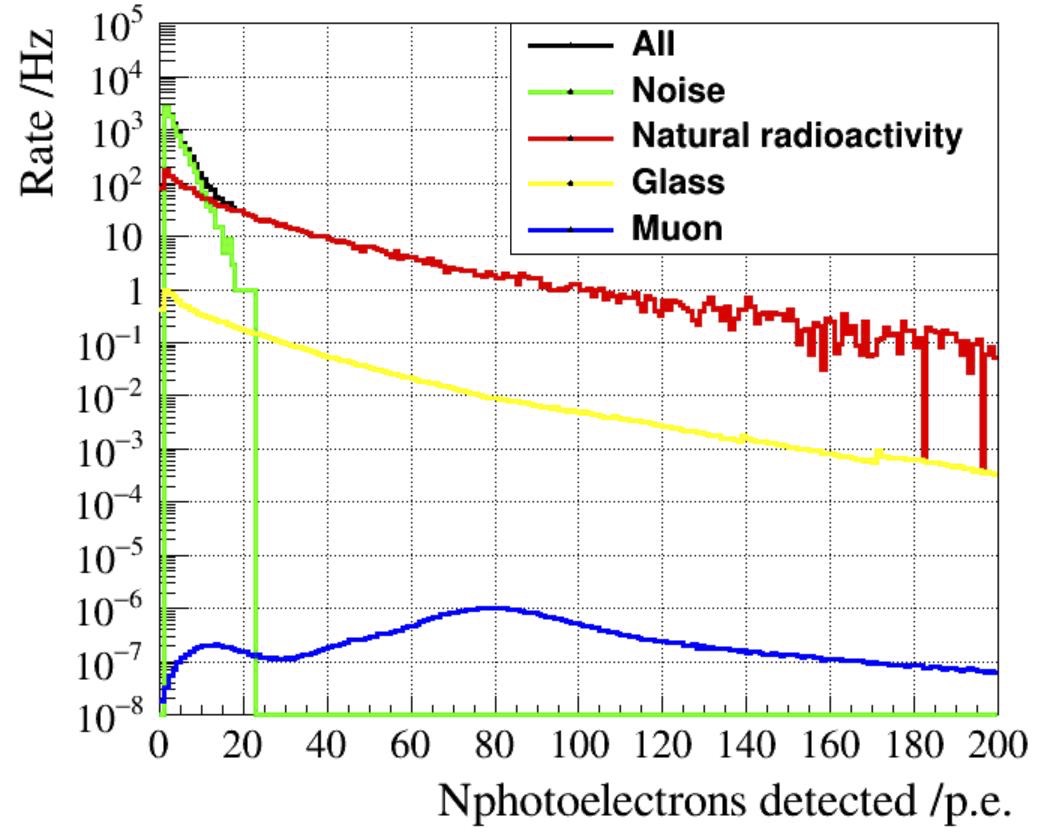}
        \label{fig:juno-nnvt}
    }
    \caption{Prediction of 20 inch HPK and NNVT PMT dark noise spectrum for JUNO underground without pure water shielding (before filling), including the contribution from thermal noise, muon (0.004 Hz/m$^2$ in the JUNO detector \cite{abusleme2021juno}), natural radioactivity of rock, and PMT glass. (a) HPK PMT and (b) NNVT PMT. Note: the black line is covered by the individual contributions on the plot.}
    \label{fig:juno-like}
\end{figure*}

\iffalse
\begin{figure*}[!ht]
	\centering
	\subfigure[]{
        \includegraphics[width=0.35\hsize]{figure/juno-hpk-muon.jpg}
        \label{fig:juno-hpk}
    }
    \quad
	\subfigure[]{
        \includegraphics[width=0.35\hsize]{figure/juno-hpk1.jpg}
        \label{fig:juno-hpk1}
    }
	\subfigure[]{
        \includegraphics[width=0.35\hsize]{figure/juno-nnvt-muon.jpg}
        \label{fig:juno-nnvt}
    }
    \quad
	\subfigure[]{
        \includegraphics[width=0.35\hsize]{figure/juno-nnvt1.jpg}
        \label{fig:juno-nnvt1}
    }    
    \caption{Prediction of 20 inch HPK/NNVT PMT dark noise spectrum for JUNO underground without pure water shielding (before filling), contains noise, muon, all of natural radioactivity and contrast of natural radioactivity of glass. (a) 20-inch HPK PMT charge spectrum with consideration of cosmic muon; (b) 20-inch NNVT PMT fitting charge spectrum with the consideration cosmic muon. The muon rate in the JUNO detector are 0.004 Hz/m$^2$\cite{abusleme2021juno}.}
    \label{fig:juno-like}
\end{figure*}
\fi

\section{Summary}
\label{1:summary}
This study focuses on investigating the occurrence of large pulses in the dark count of 20-inch PMTs attributed to natural radioactivity, excluding contributions from muons. These large pulses primarily originate from photons generated in the PMT glass by Cherenkov radiation with going through natural radioactivity from the surrounding environment. Both HPK and NNVT PMTs were analyzed in terms of their dark count rate (DCR), and the results were compared and fitted using a combination of measurements and a Geant4-based simulation. This approach enabled the determination of the expected charge spectra resulting from the passage of natural radioactivity through both types of PMTs.
The measurement of the PMT DCR spectrum plays a crucial role in providing a rough estimation of the radioactivity present in the environment. Additionally, it allows for the preliminary prediction of the 20-inch PMT spectra influenced by natural radioactivity in specific experimental environments, such as the JUNO experiment.

\acknowledgments
This work was supported partially by the National Natural Science Foundation of China (Grant No. 11875282 and 12022505), the Strategic Priority Research Program of the Chinese Academy of Sciences (Grant No. XDA10011200), and the Youth Innovation Promotion Association of CAS.

% Bibliography
 \bibliographystyle{JHEP}
 \bibliography{biblio.bib}

\providecommand{\href}[2]{#2}\begingroup\raggedright\begin{thebibliography}{10}

\bibitem{Super-Kamiokande:1998uiq}
{\scshape Super-Kamiokande} collaboration, \emph{{Measurement of the flux and zenith angle distribution of upward through going muons by Super-Kamiokande}}, \href{https://doi.org/10.1103/PhysRevLett.82.2644}{\emph{Phys. Rev. Lett.} {\bfseries 82} (1999) 2644} [\href{https://arxiv.org/abs/hep-ex/9812014}{{\ttfamily hep-ex/9812014}}].

\bibitem{PhysRevD.83.052010}
{\scshape {Super-Kamiokande Collaboration}} collaboration, \emph{{Solar neutrino results in Super-Kamiokande-III}}, \href{https://doi.org/10.1103/PhysRevD.83.052010}{\emph{Phys. Rev. D} {\bfseries 83} (2011) 052010}.

\bibitem{PhysRevLett.90.021802}
{\scshape {KamLAND Collaboration}} collaboration, \emph{{First Results from KamLAND: Evidence for Reactor Antineutrino Disappearance}}, \href{https://doi.org/"10.1103/PhysRevLett.90.021802"}{\emph{{Phys. Rev. Lett.}} {\bfseries {90}} ({2003}) {021802}}.

\bibitem{SNO}
{\scshape {SNO Collaboration}} collaboration, \emph{{Measurement of day and night neutrino energy spectra at SNO and constraints on neutrino mixing parameters}}, \href{https://doi.org/{10.1103/PhysRevLett.89.011302}}{\emph{{Phys Rev Lett.}} {\bfseries 89} (2002) 011302}.

\bibitem{PhysRevLett.110.131302}
{\scshape {IceCube Collaboration}} collaboration, \emph{{Search for Dark Matter Annihilations in the Sun with the 79-String IceCube Detector}}, \href{https://doi.org/10.1103/PhysRevLett.110.131302}{\emph{{Phys. Rev. Lett.}} {\bfseries 110} (2013) 131302}.

\bibitem{chooz}
{\scshape "Chooz Collaboration"} collaboration, \emph{{Indication of reactor new(e) disappearance in the Double Chooz experiment}}, \href{https://doi.org/"10.1103/PhysRevLett.108.131801"}{\emph{"Phys Rev Lett."} {\bfseries "108"} ("2012") "131801"}.

\bibitem{dayabay}
{\scshape Daya Bay Collaboration} collaboration, \emph{{Observation of electron-antineutrino disappearance at Daya Bay}}, \href{https://doi.org/10.1103/PhysRevLett.108.171803}{\emph{Phys Rev Lett.} {\bfseries 108} (2012) 171803}.

\bibitem{KIM201324}
S.-B.~Kim, \emph{{Observation of Reactor Electron Antineutrino Disappearance at RENO}}, \href{https://doi.org/10.1016/j.nuclphysbps.2013.03.006}{\emph{Nuclear Physics B - Proceedings Supplements} {\bfseries 235-236} (2013) 24}.

\bibitem{AugerPMT}
D.~Barnhill et~al., \emph{{Testing of photomultiplier tubes for use in the surface detector of the Pierre Auger Observatory}}, \href{https://doi.org/10.1016/j.nima.2008.01.088}{\emph{Nucl. Instrum. Meth. A} {\bfseries 591} (2008) 453}.

\bibitem{GE2016175}
M.~Ge, L.~Zhang, Y.~Chen, Z.~Cao, S.~Zhang, C.~Wang et~al., \emph{Photomultiplier tube selection for the wide field of view cherenkov/fluorescence telescope array of the large high altitude air shower observatory}, \href{https://doi.org/10.1016/j.nima.2016.02.093}{\emph{Nuclear Instruments and Methods in Physics Research Section A} {\bfseries 819} (2016) 175}.

\bibitem{BorexinoPMT}
G.~Ranucci et~al., \emph{{Characterization and magnetic shielding of the large cathode area PMTs used for the light detection system of the prototype of the solar neutrino experiment Borexino}}, \href{https://doi.org/10.1016/0168-9002(93)91156-H}{\emph{Nucl. Instrum. Meth. A} {\bfseries 337} (1993) 211}.

\bibitem{DayabayPMT}
D.~Liu, \emph{{PMT evaluation for the Daya Bay neutrino experiment}},  in \emph{2008 {IEEE} {Nuclear Science Symposium} and {Medical Imaging Conference} and 16th {International} {Workshop} on Room-Temperature Semiconductor {X-Ray} and {Gamma-Ray} Detectors}, pp.~3133--3139, 2008, \href{https://doi.org/10.1109/NSSMIC.2008.4775017}{DOI}.

\bibitem{ChoozPMT}
A.~Baldini et~al., \emph{{The photomultiplier test facility for the reactor neutrino oscillation experiment CHOOZ and the measurements of 250 8-in.\,EMI 9356KA B53 photomultipliers}}, \href{https://doi.org/https://doi.org/10.1016/0168-9002(95)01236-2}{\emph{Nucl. Instrum. Meth. A} {\bfseries 372} (1996) 207}.

\bibitem{HKPMT}
C.~Bronner et~al., \emph{{Development and performance of the 20'' PMT for Hyper-Kamiokande}}, \href{https://doi.org/10.1088/1742-6596/1468/1/012237}{\emph{J. Phys. Conf. Ser.} {\bfseries 1468} (2020) 012237}.

\bibitem{JUNO3inchPMT}
C.~Cao et~al., \emph{{Mass production and characterization of 3-inch PMTs for the JUNO experiment}}, \href{https://doi.org/10.1016/j.nima.2021.165347}{\emph{Nucl. Instrum. Meth. A} {\bfseries 1005} (2021) 165347} [\href{https://arxiv.org/abs/2102.11538}{{\ttfamily 2102.11538}}].

\bibitem{JUNOPMTinstr}
Z.~Qin, \emph{{Status of the 20-in. PMT Instrumentation for the JUNO Experiment}},  in \emph{Proceedings of International Conference on Technology and Instrumentation in Particle Physics 2017}, pp.~285--293, Springer Singapore, 2018.

\bibitem{KM3NeTPMT}
{\scshape KM3NeT} collaboration, \emph{{Development and performances of a high statistics PMT test facility}}, \href{https://doi.org/10.1051/epjconf/201611606010}{\emph{EPJ Web Conf.} {\bfseries 116} (2016) 06010}.

\bibitem{JUNOPMTflasher}
A.~Yang et~al., \emph{{Study and removal of the flash from the HV divider of the 20-inch PMT for JUNO}}, \href{https://doi.org/10.1088/1748-0221/15/04/T04006}{\emph{JINST} {\bfseries 15} (2020) T04006}.

\bibitem{MCPPMT2018}
S.~Qian et~al., \emph{{The improvement of 20'' MCP-PMT for neutrino detection}}, \href{https://doi.org/10.22323/1.340.0662}{\emph{PoS} {\bfseries ICHEP2018} (2019) 662}.

\bibitem{YWang_newMCP}
Y.~Wang et~al., \emph{{A new design of large area MCP-PMT for the next generation neutrino experiment}}, \href{https://doi.org/10.1016/j.nima.2011.12.085}{\emph{Nucl. Instrum. Meth. A} {\bfseries 695} (2012) 113}.

\bibitem{wavesamplingPMT}
S.~Yin et~al., \emph{{A novel PMT test system based on waveform sampling}}, \href{https://doi.org/10.1088/1748-0221/13/01/T01005}{\emph{JINST} {\bfseries 13} (2018) T01005}.

\bibitem{waveAnalysisHaiqiong}
H.Q.~Zhang et~al., \emph{{Comparison on PMT Waveform Reconstructions with JUNO Prototype}}, \href{https://doi.org/10.1088/1748-0221/14/08/T08002}{\emph{JINST} {\bfseries 14} (2019) T08002} [\href{https://arxiv.org/abs/1905.03648}{{\ttfamily 1905.03648}}].

\bibitem{Abe_2016}
Y.A.~et~al., \emph{Characterization of the spontaneous light emission of the {PMTs} used in the double chooz experiment}, \href{https://doi.org/10.1088/1748-0221/11/08/p08001}{\emph{Journal of Instrumentation} {\bfseries 11} (2016) P08001}.

\bibitem{DWYER201330}
D.~Dwyer, \emph{Improved measurement of electron-antineutrino disappearance at daya bay}, \href{https://doi.org/10.1016/j.nuclphysbps.2013.03.007}{\emph{Nuclear Physics B} {\bfseries 235-236} (2013) 30}.

\bibitem{IceCube-inproceedings}
{\scshape IceCube} collaboration, \emph{{Characterisation of Two PMT Models for the IceCube Upgrade mDOM}}, \href{https://doi.org/10.22323/1.358.1022}{\emph{PoS} {\bfseries ICRC2019} (2020) 1022} [\href{https://arxiv.org/abs/1908.08446}{{\ttfamily 1908.08446}}].

\bibitem{JANG2014145}
J.~Jang, \emph{A precise measurement of reactor antineutrino at reno}, \href{https://doi.org/10.1016/j.nds.2014.07.030}{\emph{Nuclear Data Sheets} {\bfseries 120} (2014) 145}.

\bibitem{PMTmuon2007}
L.V.~"W.~Raposo, M.~Vaz, \emph{"measurements of signals from muons crossing the hamamatsu r5912 pmt enclosure vertically and horizontally"}, {\emph{online} ("2007") }.

\bibitem{BAYAT20141}
E.~Bayat, V.~Doust-Mohammadi, P.~Ghorbani, N.~Ghal-Eh and R.~Mohammadi, \emph{Scintillation of xp2020 pmt glass window}, \href{https://doi.org/https://doi.org/10.1016/j.radphyschem.2014.04.009}{\emph{Radiation Physics and Chemistry} {\bfseries 102} (2014) 1}.

\bibitem{Zhang_2022}
Y.~Zhang, Z.~Wang, M.~Li, Y.~Zhang, Y.~Wang, Z.~Peng et~al., \emph{Study of 20-inch pmts dark count generated large pulses}, \href{https://doi.org/10.1088/1748-0221/17/10/P10048}{\emph{Journal of Instrumentation} {\bfseries 17} (2022) P10048}.

\bibitem{glass-2015}
L.~Li-Wan et~al., \emph{Scintillation properties of ce3+ doped sio2-al2o3-gd2o3 glass}, \href{https://doi.org/10.7498/aps.64.167802}{\emph{Acta Physica Sinica} {\bfseries 64} (2015) 167802}.

\bibitem{TANG2022112585-2022}
G.~Tang, Z.~Hua, S.~Qian, X.~Sun, H.~Ban, H.~Cai et~al., \emph{Optical and scintillation properties of aluminoborosilicate glass}, \href{https://doi.org/https://doi.org/10.1016/j.optmat.2022.112585}{\emph{Optical Materials} {\bfseries 130} (2022) 112585}.

\bibitem{AMELINA2022121393-2022}
A.~Amelina, A.~Mikhlin, S.~Belus, A.~Bondarev, A.~Borisevich, D.~Kuznetsova et~al., \emph{(gd,ce)2o3-al2o3-sio2 scintillation glass}, \href{https://doi.org/https://doi.org/10.1016/j.jnoncrysol.2021.121393}{\emph{Journal of Non-Crystalline Solids} {\bfseries 580} (2022) 121393}.

\bibitem{7104168}
M.L.~Ahnen, J.~Hose, U.~Menzel and R.~Mirzoyan, \emph{Light induced afterpulses in photomultipliers}, \href{https://doi.org/10.1109/TNS.2015.2416775}{\emph{IEEE Transactions on Nuclear Science} {\bfseries 62} (2015) 1313}.

\bibitem{CAO201662-dayabay-flasher}
J.~Cao and K.-B.~Luk, \emph{An overview of the daya bay reactor neutrino experiment}, \href{https://doi.org/https://doi.org/10.1016/j.nuclphysb.2016.04.034}{\emph{Nuclear Physics B} {\bfseries 908} (2016) 62}.

\bibitem{Qian_2020}
S.~Qian, F.~Gao, L.~Ma, Y.~Zhu, S.~Chen and P.~Chen, \emph{The study on the 20 inch {PMT} flasher signal}, \href{https://doi.org/10.1088/1748-0221/15/06/t06008}{\emph{Journal of Instrumentation} {\bfseries 15} (2020) T06008}.

\bibitem{MIRZOYAN199774-AP}
R.~Mirzoyan, E.~Lorenz, D.~Petry and C.~Prosch, \emph{On the influence of afterpulsing in pmts on the trigger threshold of multichannel light detectors in self-trigger mode}, \href{https://doi.org/https://doi.org/10.1016/S0168-9002(96)00964-3}{\emph{Nuclear Instruments and Methods in Physics Research Section A: Accelerators, Spectrometers, Detectors and Associated Equipment} {\bfseries 387} (1997) 74}.

\bibitem{JUNO-PMT-AP}
R.~Zhao, N.~Anfimov, Y.~Chen and et~al., \emph{Afterpulse measurement of juno 20-inch pmts}, \href{https://doi.org/https://doi.org/10.1007/s41365-022-01162-3}{\emph{NUCL SCI TECH} {\bfseries 34} (2023) 12}.

\bibitem{Yang_2020}
A.~Yang, Z.~Qin, Z.~Wang, H.~Chen, W.~Wei, F.~Luo et~al., \emph{Study and removal of the flash from the {HV} divider of the 20-inch {PMT} for {JUNO}}, \href{https://doi.org/10.1088/1748-0221/15/04/t04006}{\emph{Journal of Instrumentation} {\bfseries 15} (2020) T04006}.

\bibitem{JUNOCDR}
{\scshape JUNO} collaboration, \emph{{JUNO Conceptual Design Report}}, {\emph{arXiv e-prints} (2015) } [\href{https://arxiv.org/abs/1508.07166}{{\ttfamily 1508.07166}}].

\bibitem{JUNOphysics}
{\scshape JUNO} collaboration, \emph{{Neutrino Physics with JUNO}}, \href{https://doi.org/10.1088/0954-3899/43/3/030401}{\emph{J. Phys. G} {\bfseries 43} (2016) 030401} [\href{https://arxiv.org/abs/1507.05613}{{\ttfamily 1507.05613}}].

\bibitem{JUNOdetector}
J.~collaboration, \emph{{JUNO} physics and detector}, \href{https://doi.org/https://doi.org/10.1016/j.ppnp.2021.103927}{\emph{Progress in Particle and Nuclear Physics} {\bfseries 123} (2022) 103927}.

\bibitem{HPK-R12860}
H.P.~K.K., \emph{{R12860 datasheet}},  2019.

\bibitem{NNVT-GDB6201-note}
N.N.V.T.~Ltd., \emph{{Specification for GDB-6201 microchannel plate type photomultiplier PMT (in Chinese)}},  2020.

\bibitem{HamManual}
{Hamamatsu Photonics K.K}, \emph{{Photomultiplier Tubes - Basics and Applications}}, Hamamatsu Photonics K.K., {3rd}~ed. (2007).

\bibitem{POLYAKOV201369}
S.V.~Polyakov, \emph{Chapter 3 - photomultiplier tubes},  in \emph{Single-Photon Generation and Detection}, A.M.~et~al., ed., vol.~45 of \emph{Experimental Methods in the Physical Sciences}, pp.~69--82, Academic Press (2013), \href{https://doi.org/10.1016/B978-0-12-387695-9.00003-2}{DOI}.

\bibitem{zhang2022study}
Y.~Zhang, Z.~Wang, M.~Li, Y.~Zhang, Y.~Wang, Z.~Peng et~al., \emph{Study of 20-inch pmts dark count generated large pulses}, {\emph{Journal of Instrumentation} {\bfseries 17} (2022) P10048}.

\bibitem{JUNOPMTgain}
H.Q.~Zhang et~al., \emph{{Gain and charge response of 20'' MCP and dynode PMTs}}, \href{https://doi.org/10.1088/1748-0221/16/08/T08009}{\emph{JINST} {\bfseries 16} (2021) T08009} [\href{https://arxiv.org/abs/2103.14822}{{\ttfamily 2103.14822}}].

\bibitem{MCPPMTTTSsen-reflection}
Q.~Sen, ``{The Large Area MCP-PMT for Neutrino Detector}.''

\bibitem{PMTgainmodel1994}
E.H.~Bellamy et~al., \emph{{Absolute calibration and monitoring of a spectrometric channel using a photomultiplier}}, \href{https://doi.org/10.1016/0168-9002(94)90183-X}{\emph{Nucl. Instrum. Meth. A} {\bfseries 339} (1994) 468}.

\bibitem{PMTgainmodel2017}
R.~Saldanha et~al., \emph{{Model Independent Approach to the Single Photoelectron Calibration of Photomultiplier Tubes}}, \href{https://doi.org/10.1016/j.nima.2017.02.086}{\emph{Nucl. Instrum. Meth. A} {\bfseries 863} (2017) 35} [\href{https://arxiv.org/abs/1602.03150}{{\ttfamily 1602.03150}}].

\bibitem{TAKAHASHI20181-PMT-gainmodel}
M.~Takahashi, Y.~Inome, S.~Yoshii, A.~Bamba, S.~Gunji, D.~Hadasch et~al., \emph{A technique for estimating the absolute gain of a photomultiplier tube}, \href{https://doi.org/https://doi.org/10.1016/j.nima.2018.03.034}{\emph{Nuclear Instruments and Methods in Physics Research Section A: Accelerators, Spectrometers, Detectors and Associated Equipment} {\bfseries 894} (2018) 1}.

\bibitem{Luo_2019}
F.L.~et~al., \emph{A study of the new hemispherical 9-inch {PMT}}, \href{https://doi.org/10.1088/1748-0221/14/02/t02004}{\emph{Journal of Instrumentation} {\bfseries 14} (2019) T02004}.

\bibitem{geant4}
G.W.G..~Coordinators, \emph{{Geant4}},  2021.

\bibitem{ZHANG201867-MCP-glass}
X.~Zhang, J.~Zhao, S.~Liu, S.~Niu, X.~Han, L.~Wen et~al., \emph{Study on the large area mcp-pmt glass radioactivity reduction}, \href{https://doi.org/https://doi.org/10.1016/j.nima.2018.05.008}{\emph{Nuclear Instruments and Methods in Physics Research Section A: Accelerators, Spectrometers, Detectors and Associated Equipment} {\bfseries 898} (2018) 67}.

\bibitem{Guan2015APO-muonflux}
M.~Guan, M.~Chu, J.~Cao, K.B.~Luk and C.~Yang, \emph{A parametrization of the cosmic-ray muon flux at sea-level}, {\emph{arXiv: High Energy Physics - Experiment} (2015) }.

\bibitem{PhysRevD.58.054001-muonflux}
E.V.~Bugaev, A.~Misaki, V.A.~Naumov, T.S.~Sinegovskaya, S.I.~Sinegovsky and N.~Takahashi, \emph{Atmospheric muon flux at sea level, underground, and underwater}, \href{https://doi.org/10.1103/PhysRevD.58.054001}{\emph{Phys. Rev. D} {\bfseries 58} (1998) 054001}.

\bibitem{armengaud2017performance}
E.~Armengaud, Q.~Arnaud, C.~Augier, A.~Beno{\^\i}t, L.~Berg{\'e}, T.~Bergmann et~al., \emph{Performance of the edelweiss-iii experiment for direct dark matter searches}, {\emph{Journal of Instrumentation} {\bfseries 12} (2017) P08010}.

\bibitem{abusleme2021juno}
A.~Abusleme, T.~Adam, S.~Ahmad, R.~Ahmed, S.~Aiello, M.~Akram et~al., \emph{Juno physics and detector}, {\emph{arXiv preprint arXiv:2104.02565} (2021) }.

\bibitem{junocollaboration2023juno}
{\scshape JUNO} collaboration, \emph{{The JUNO experiment Top Tracker}}, {\emph{arXiv preprint arXiv:2303.05172} (2023) } [\href{https://arxiv.org/abs/2303.05172}{{\ttfamily 2303.05172}}].

\end{thebibliography}\endgroup
\end{document}